\documentclass[pra,showpacs,amsmath,amssymb,aps,superscriptaddress,twocolumn]{revtex4-2}
\usepackage{graphicx}
\usepackage{amsthm}
\usepackage{amssymb}
\usepackage{amsmath}
\usepackage{braket}
\usepackage{bm}
\usepackage{here}
\usepackage{hyperref}
\newtheorem{theorem}{Theorem}
\newtheorem{lemma}{Lemma}
\newtheorem{prop}{Proposition}

\begin{document}
\title{Refined security proof of the round-robin differential phase shift quantum key distribution and its improved performance in the finite-sized case}
\author{Takaya Matsuura}
 \email{matsuura@qi.t.u-tokyo.ac.jp}
 \affiliation{Department of Applied Physics, Graduate School of Engineering, The University of Tokyo, 7-3-1 Hongo, Bunkyo-ku, Tokyo 113-8656, Japan} 
\author{Toshihiko Sasaki}
 \affiliation{Department of Applied Physics, Graduate School of Engineering, The University of Tokyo, 7-3-1 Hongo, Bunkyo-ku, Tokyo 113-8656, Japan} 
 \affiliation{Photon Science Center, Graduate School of Engineering, The University of Tokyo, 7-3-1 Hongo, Bunkyo-ku, Tokyo 113-8656, Japan} 
\author{Masato Koashi}
 \affiliation{Department of Applied Physics, Graduate School of Engineering, The University of Tokyo, 7-3-1 Hongo, Bunkyo-ku, Tokyo 113-8656, Japan} 
 \affiliation{Photon Science Center, Graduate School of Engineering, The University of Tokyo, 7-3-1 Hongo, Bunkyo-ku, Tokyo 113-8656, Japan}
   
 \date{\today}

\begin{abstract}
Among many quantum key distribution (QKD) protocols, the round-robin differential phase shift (RRDPS) protocol is unique in that it can upper-bound the amount of the information leakage without monitoring the signal disturbance.   
To expedite implementation of the protocol, however, the number of pulses forming a single block should be kept small, which significantly decreases the key rates in the original security proof.
In the present paper, we refine the security proof of the RRDPS protocol in the finite-sized regime and achieve a tighter estimation for the information leakage without changing the original experimental setups.  As a consequence, we obtain better key rates in both asymptotic and finite-sized cases while keeping the preferable features of the protocol, such as omission of phase randomization.  
\end{abstract}
\maketitle

\section{Introduction} 
 
One of the most important implications of the quantum information theory is that information-theoretically secure communication is possible by the quantum key distribution (QKD).  After the first proposal of the BB84 protocol \cite{Bennett1984}, many researches have been made in the field.  In recent years, the real world implementation of the QKD is attracting much attention.  For the real world implementation, we need careful consideration about the finite-sized effect of the key and the imperfections of the experimental devices because communications in the real world are often done in limited time and with imperfect devices.  The finite-sized key rate of the QKD protocol is especially important when we consider the communication between the ground and the satellite \cite{Vallone2015,Liao2017} for which the communication time is limited and therefore only a small number of bits can be sent at a time.

The round-robin differential phase shift (RRDPS) protocol \cite{Sasaki2014} is a QKD protocol which has a special property that the required amount of privacy amplification is determined only by the protocol parameters and independent of the bit error rates.  Due to this property, the protocol is expected to be able to generate the key even when the number of communication rounds is small, because it does not suffer from the convoluted statistical estimation of the information leakage. 
The protocol can be implemented with a light source producing a coherent laser pulse train at the sender, and a variable-delay interferometer followed by photon detection at the receiver.  A number of experimental demonstrations have already been made \cite{Takesue2015,Wang2015,Guan2015,Li2016}.  Especially, the apparatus for the sender can be made very simple with only binary phase modulation, and the security can be proved without phase randomization of the optical pulses.
Fewer assumptions on the light source in the RRDPS protocol also lead to the robustness against the source imperfection \cite{Mizutani2015}.  

On the other hand, the RRDPS protocol also has a few undesirable features.  The protocol assumes a variable delay interferometer which should be switched among $L$ different delays actively or passively for each pulse block.  Implementing such an interferometer is costly especially for large $L$.  Furthermore, the asymptotic key rate of the RRDPS protocol even with relatively large block size ($L \sim 128$) is about one-tenth of that of the decoy BB84 protocol \cite{Hwang2003}, which is a widely used and the most studied practical QKD protocol.  The key rate gets even worse when we decrease $L$ to simplify the implementation.  Therefore, it is desired to improve the key rate of the RRDPS protocol especially for relatively small $L$.  There have been intensive researches to mitigate or to get over these problems both in theory \cite{Yin2016,Zhang2017,Sasaki2017,Hatakeyama2017,WangLe2017,Liu2017,Leermakers2017,Yin2018} and experiment \cite{Guan2015,Bouchard2018}.
        
Very recently, Yin {\it et al}.\ shows that by directly evaluating Eve's collective attacks, one can improve the key rate of the RRDPS protocol with block-wise phase randomization without any change in the protocol \cite{Yin2018}.  It also implies that we can decrease $L$ to achieve the same key rate.
Unfortunately, the analysis in \cite{Yin2018} cannot directly be extended to the finite-sized case, and thus its usage is limited.  
        
In this paper, we refine the security proof of the RRDPS protocol with a different approach and obtain better key rates in both asymptotic and finite-sized case without block-wise phase randomization.  The main idea of our analysis is to utilize the information disregarded in the original security proof, which leads to a tighter estimation for the amount of the information leakage without the aid of the block-wise phase randomization.  Our analysis developed here is based on the technique used in the security proof of the differential quadrature phase shift protocol \cite{Kawakami2016}, and it may be applicable to other high dimensional QKD protocols including other DPS-type protocols.  The obtained key rate in the asymptotic limit with our analysis is almost the same as that in \cite{Yin2018}, but we do not require the block-wise phase randomization, and we can also explicitly give the key rate formula in the finite-sized case.  Furthermore, we show that the RRDPS protocol outperforms decoy BB84 protocol when the number of communication rounds is small.  

The paper is organized as follows.  In Section 2, we develop the refined security proof of the RRDPS protocol, which is the main part of this paper.  We give the definition of the protocol and subsequently construct a compatible virtual protocol which includes a crucial difference from the original one.  We further introduce another auxiliary protocol which reproduces the statistics of the phase errors in the virtual protocol, and by analyzing it, we derive the main theorem, which gives the required amount of the privacy amplification.  In Section 3, we numerically simulate the key rates of the RRDPS protocol with our refined analysis in both asymptotic and finite-sized case, illustrating how we determine the parameters which appear in the key rate formula.  Finally, in Section 4, we wrap up our analysis, discuss the comparison between the techniques developed here and the existing ones, and refer to some remaining problems.

\section{Security proof}

In what follows, $h(x):=-x\log x - (1 - x)\log (1 - x)$ denotes binary entropy function, $H(X|Y)_P$ denotes the conditional entropy with the joint probability distribution $P$, and $D(P \| Q)$ denotes the Kullback-Leibler divergence.  $\mathbb{E}_{X\sim P}[f(X)]$ denotes the expectation value of $f(X)$ when the random variable $X$ obeys the probability distribution $P$.  $\| \rho - \sigma \|_1 =  \mathrm{Tr}|\rho - \sigma| $ is the trace norm distance and $F(\rho,\sigma) = \|\sqrt{\rho} \sqrt{\sigma}\|_1^2$ is the fidelity between the density matrices $\rho$ and $\sigma$.  We call $\{\ket{0},\ket{1}\}$ as the bit basis of the qubit, and $\{\ket{0_X}=(\ket{0} + \ket{1})/{\sqrt{2}},\ket{1_X} = (\ket{0} - \ket{1})/{\sqrt{2}}\}$ as the phase basis.  The controlled-NOT (CNOT) operation between control qubit 1 and target qubit 2 is defined as $\ket{0}\bra{0}_1 \otimes I_2 + \ket{1}\bra{1}_1 \otimes X_2 = I_1 \otimes \ket{0_X}\bra{0_X}_2 +  Z_1 \otimes \ket{1_X}\bra{1_X}_2$, where $I$ is the identity operator, $X=\ket{0}\bra{1} + \ket{1}\bra{0}$, and $Z=\ket{0_X}\bra{1_X} + \ket{1_X}\bra{0_X}$.  $\bm{I}_{N}$ denotes $N \times N$ identity matrix, and $\oplus$ denotes the summation modulo $2$.  The base of the logarithm is taken to be $2$.

\subsection{The definition of the protocol}

We first give the actual procedure of the RRDPS protocol \cite{Sasaki2014} and the assumptions for the analysis in this paper.  \\

\noindent{\bf Setups and assumptions:}
The sender Alice has a $0/\pi$ phase modulator and an i.i.d.\ source of weak coherent optical pulses.  The quantum state of each optical pulse is represented by a density operator $\sigma$, which has no correlation with any other system.  The probability that the source emits odd numbers of photons is upper-bounded by a known parameter $p_\mathrm{src}$ (e.g.\ $p_\mathrm{src}=1 - \bra{0}\sigma\ket{0}$).  Bob has a variable delay interferometer whose delay can be switched according to randomly generated numbers.  The photon detector can distinguish zero, one, and two or more photons.  The inefficiency and the dark counting of the photon detectors can be included in the channel loss.  They share a public channel for announcement as well as a quantum channel.  The eavesdropper Eve can perform arbitrary attacks allowed in the law of quantum mechanics on the quantum channel and listen to all the announcements of Alice and Bob made over the public channel.\\

\begin{figure}[t]
    \centering
    \includegraphics[width=86mm]{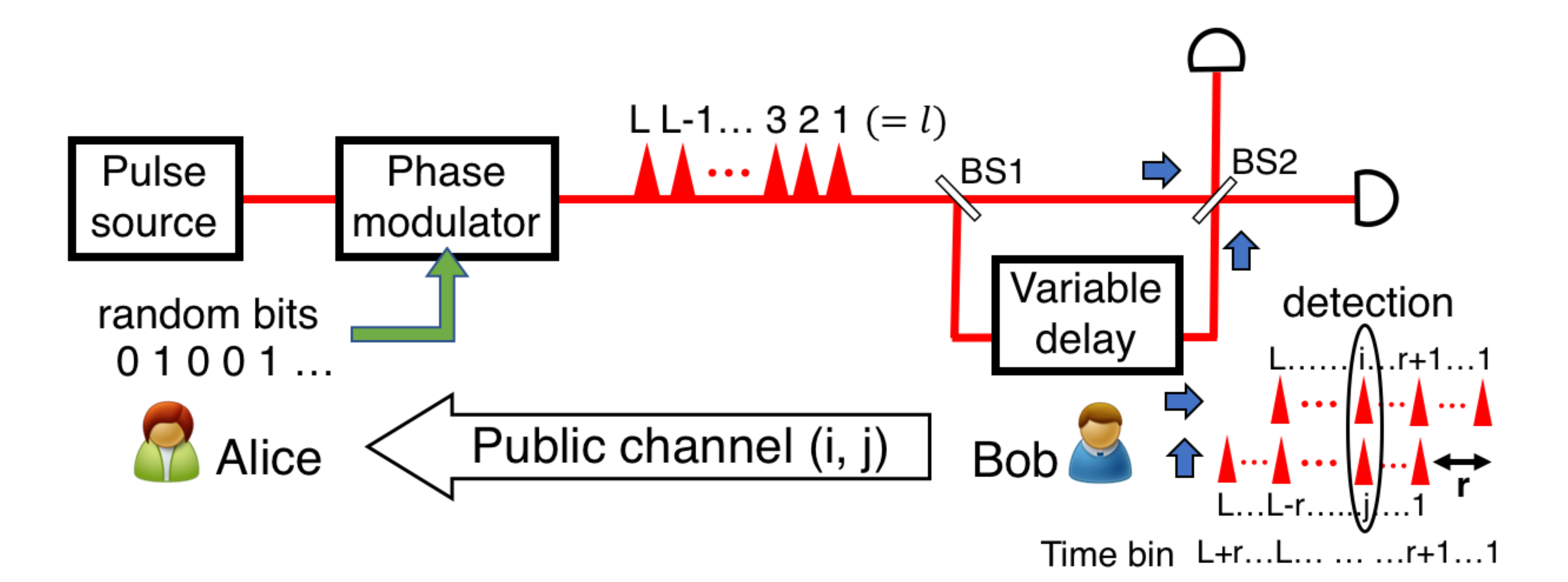}
    \caption{Schematics of the RRDPS protocol}
    \label{fig:RRDPS} 
\end{figure}

\noindent{\bf Protocol 1 (actual protocol):}

Before the commencement of the protocol, Alice and Bob agree on constants $L$ and $N_{\mathrm{em}}$ as well as a function $N_{\mathrm{fin}}(N)$ and probabilities $p'(C|N,N_{\mathrm{fin}})$ over full-rank $N\times N_{\mathrm{fin}}$ binary matrices $C$.
\begin{enumerate}
    \renewcommand{\labelenumi}{(\roman{enumi})}
    \item Alice and Bob repeat the following procedures for $N_{\mathrm{em}}$ rounds.
    \begin{itemize}
        \item Alice generates a sequence of random bits $s^{(1)},...,s^{(L)}$, and encode them to $L$ optical pulses 
        by modulating the optical phase of the $l$-th pulse with $\mathrm{e}^{\pi \mathrm{i} s^{(l)}}$ ($l=1,...,L$).  She sends Bob the $L$ optical pulses through the quantum channel.    
        \item Bob randomly selects the delay $r \in \{1,...,L-1\}$ and feeds the received $L$ pulses to the delayed interferometer as shown in Figure \ref{fig:RRDPS}.  He detects photons with the two detectors at time bins $1$ through $L + r$.
        \begin{itemize}
            \item If Bob detects only one photon from the $(r+1)$-th to the $L$-th time bin, and observes no detection at the other bins, he records a sifted key bit $z_B\in\{0,1\}$ according to which photon detector has reported the detection.  He also records the unordered pair $(i,j)$, which are the positions of the pulse pair arriving at the detected time bin ($i,j\in\{1,...,L\},\ |i-j|=r$).  He announces ``success'', and Alice records her random bit sequence $(s^{(1)},...,s^{(L)})$.  [Success round]   
            \item If the above condition is not satisfied, Bob announces ``failure'' and Alice discards her random bits.  [Failure round] 
        \end{itemize} 
    \end{itemize}
    \item Let $N$ be the number of the success rounds.  By proper indexing, Alice's records are represented by $(s^{(1)}_{1},...,s^{(L)}_{1}),...,(s^{(1)}_{N},...,s^{(L)}_{N})$, and Bob's sifted key by $\bm{z}_B = ({z_{B}}_{1}\ \cdots \ {z_{B}}_{N})$ and his unordered pairs by $(i_{1},j_{1}),...,(i_{N},j_{N})$.  
    \item Bob announces the sequence of the unordered pairs $(i_{1},j_{1}), ...,(i_{N},j_{N})$.  
    \item Alice defines her sifted key $\bm{z}_A = ({z_{A}}_{1}\ \cdots \ {z_{A}}_{N})$ by ${z_{A}}_{k} := s^{(i_k)}_{k} \oplus s^{(j_k)}_{k}$ for $k=1,...,N$.
    \item (Bit error correction)  Alice chooses and announces a bit error correcting code.  She calculates the $N_{EC} $-bit syndrome for $\bm{z}_A$ and encrypts it by consuming $N_{EC} $ bits of the pre-shared secret key before she sends it to Bob.  With the syndrome, Bob performs bit error correction on his sifted key $\bm{z}_B$ and obtains the reconciled key $\bm{z}_{B}^{\mathrm{rec}}$ of $N$ bits.  
    \item (Privacy amplification)  Let $N_{\mathrm{fin}}:=N_{\mathrm{fin}}(N)$.  Alice draws a full-rank $N\times N_{\mathrm{fin}}$ binary matrix $C_{PA}$ with the probability $p'(C_{PA}|N,N_{\mathrm{fin}})$ and announces it.  Alice and Bob computes the final keys as $\bm{z}_A^{\text{fin}} = \bm{z}_A C_{PA}$ and $\bm{z}_B^{\text{fin}} = \bm{z}_B^{\mathrm{rec}} C_{PA}$, respectively.   
\end{enumerate} 

For simplicity, we omitted the bit error sampling rounds in the above protocol.  In order to estimate an upper-bound on the bit error rate $e_{\mathrm{bit}}$, Alice randomly inserts $N_{\mathrm{smp}}$ sampling rounds among $N_{\mathrm{em}}$ rounds, and according to $e_{\mathrm{bit}}$, she decides whether she aborts the protocol or not.  
Here we assume that $N_{\mathrm{smp}}$ is negligibly small compared to $N_{\mathrm{em}}$.  The required amount of the error syndrome Alice sends to Bob in the bit error correction, $N_{EC} $, depends on the error correction method; here we assume $N_{EC} = Nf_{EC}h(e_{\mathrm{bit}})$, where $f_{EC}$ is an error correction efficiency to satisfy the required correctness.  The net key gain per pulse of the protocol is therefore given by $(N_{\mathrm{fin}} - N_{EC}) / (N_{\mathrm{em}}L) = (N_{\mathrm{fin}} - Nf_{EC}h(e_{\mathrm{bit}})) / (N_{\mathrm{em}}L)$.

We evaluate the secrecy of Protocol 1 by the $\varepsilon_{\mathrm{sec}}$-secrecy condition for Alice's final key defined as
\begin{equation}
    \frac{1}{2} \sum_{N_{\mathrm{fin}}\geq 1}\mathrm{Pr}(N_{\mathrm{fin}}) \big\| \rho_{AE|N_{\mathrm{fin}}}^{\mathrm{fin}} - \rho_{AE|N_{\mathrm{fin}}}^{\mathrm{ideal}} \big\|_1 \leq \varepsilon_{\mathrm{sec}}.
    \label{eq:secrecy}
\end{equation}
Here $\mathrm{Pr}(N_{\mathrm{fin}})$ is the probability of obtaining $N_{\mathrm{fin}}$, where aborting the protocol is interpreted as $N_{\mathrm{fin}} = 0$.  The density operator $\rho_{AE|N_{\mathrm{fin}}}^{\mathrm{fin}}$ represents the state of Alice's final key and Eve's quantum system, which takes the form of
\begin{equation}
    \rho_{AE|N_{\mathrm{fin}}}^{\mathrm{fin}} = \sum_{\bm{z}_A^\mathrm{fin}\in\{0,1\}^{N_{\mathrm{fin}}}} \mathrm{Pr}(\bm{z}_A^{\text{fin}}) \ket{\bm{z}_A^{\text{fin}}} \bra{\bm{z}_A^{\text{fin} }}_A \otimes \rho_{E|N_{\mathrm{fin}}}(\bm{z}_A^{\text{fin}} ).
        \label{eq:final_state}
\end{equation}
The ideal state $\rho_{AE|N_{\mathrm{fin}}}^{\mathrm{ideal}}$ is defined as 
\begin{equation}
    \begin{aligned}
    &\rho_{AE|N_{\mathrm{fin}}}^{\mathrm{ideal}} := \left(\sum_{\bm{z}_A^\mathrm{fin}\in\{0,1\}^{N_{\mathrm{fin}}}} \frac{1}{2^{N_{\mathrm{fin}}}} \ket{\bm{z}_A^\mathrm{fin}} \bra{\bm{z}_A^\mathrm{fin}}_A \right) \\
    & \hspace{5cm} \otimes \mathrm{Tr}_{A}\left(\rho_{AE|N_{\mathrm{fin}}}^{\mathrm{fin}}\right).
    \end{aligned}
    \label{eq:ideal_state}
\end{equation}

\subsection{The reduction of the protocol} \label{sec:reduction}
\begin{figure}[t]
    \centering
    \includegraphics[width=86mm]{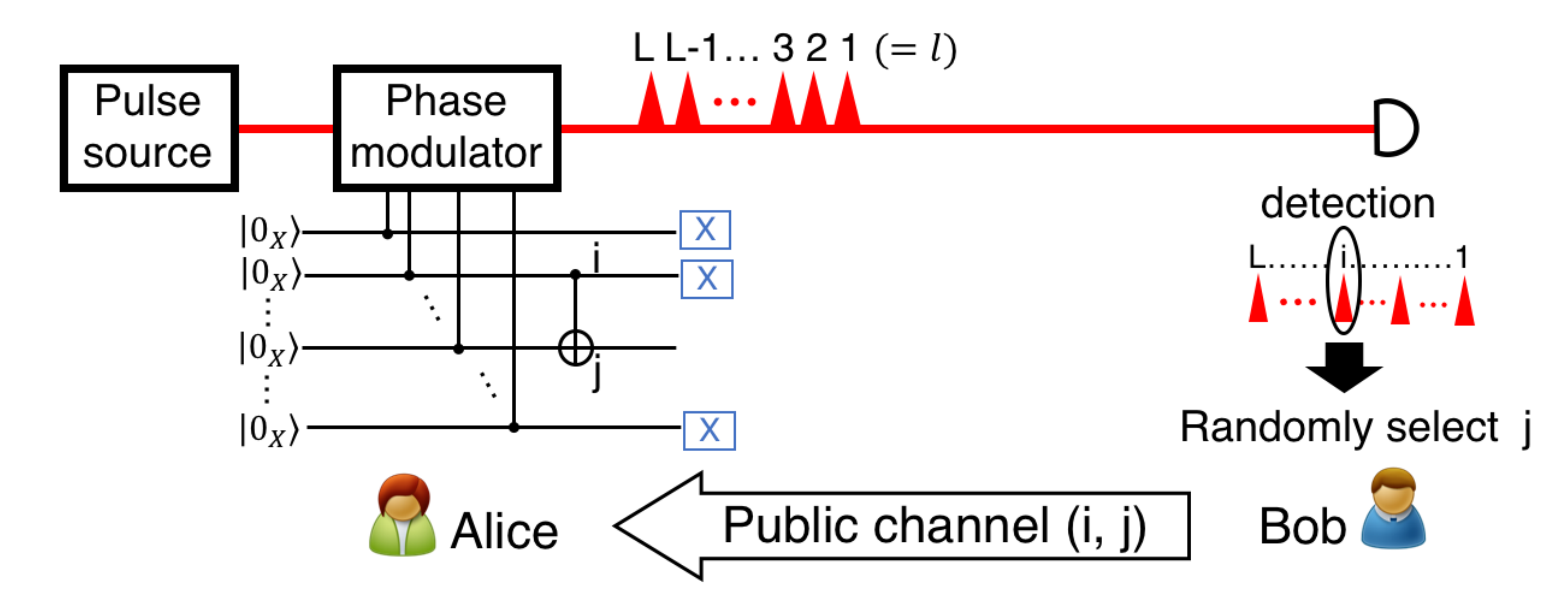}
    \caption{Virtual protocol of the RRDPS.  In contrast to the original security proof, here we assume that Alice measures all but $j$-th of the $L$ qubits in the phase basis.}
    \label{fig:RRDPS_virt} 
\end{figure}

We prove the secrecy condition (\ref{eq:secrecy}) of the protocol based on complementarity \cite{Koashi2009}.  In this way of the security proof, we introduce a virtual protocol (Protocol 2) in which Alice's $N_{\mathrm{fin}}$-bit final key is obtained by a bit basis measurement on $N_{\mathrm{fin}}$ register qubits.  Protocol 2 should be related to Protocol 1 such that for every attack on Protocol 1, there exists an attack on Protocol 2 resulting in the same final state ($\mathrm{Pr}(N_{\mathrm{fin}})$ and $\rho_{AE|N_{\mathrm{fin}}}^{\mathrm{fin}}$).  While the original proof \cite{Sasaki2014} followed the same technique, our construction of Protocol 2 below (see also Figure \ref{fig:RRDPS_virt}) includes a modification (shown in bold fonts) which is crucial to an improvement of the key rate.\\

\noindent{\bf Protocol 2 (virtual protocol):}

Before the commencement of the protocol, Alice and Bob agree on constants $L$ and $N_{\mathrm{em}}$ as well as functions $N_{\mathrm{fin}}(N)$, $\bm{x}^*(N,y^N,N_{\mathrm{fin}},\bm{t})$, and probabilities $p(C|N,N_{\mathrm{fin}})$ over full-rank $N\times N$ binary matrices $C$.  
\begin{enumerate}
    \renewcommand{\labelenumi}{(\roman{enumi})}
    \item Alice and Bob repeat the following procedures for $N_{\mathrm{em}}$ rounds.
    \begin{itemize}
        \item Alice prepares an $L$-qubit register $A^{(1)},...,A^{(L)}$, a reference $R := R^{(1)},..., R^{(L)}$, and $L$ optical pulses (system $1,...,L$) in the following state:
            \begin{equation}
                2^{- L/2}\bigotimes_{l=1}^{L} \sum_{s^{(l)} = 0, 1}\ket{s^{(l)}}_{A^{(l)}} \mathrm{e}^{\pi \mathrm{i} s^{(l)} \hat{n}_{l}}\ket{\Psi_{\sigma}}_{l R^{(l)}} ,
                \label{eq:initial_state}
            \end{equation}
            where $\mathrm{Tr}_{R}\left(\bigotimes_{l=1}^{L}\ket{\Psi_{\sigma}}\bra{\Psi_{\sigma}}_{l R^{(l)}}\right) = \sigma \otimes \cdots \otimes \sigma$, and $\hat{n}_l$ is the photon number operator for the $l$-th pulse.
            She sends Bob the $L$ optical pulses through the quantum channel. 
        \item Bob measures the photon number of each of the received pulses.  He also generates a uniformly random binary number $q$.  
            \begin{itemize}
                \item If Bob detects only one photon in the block and the generated random number $q$ is $0$, he announces ``success'' and Alice keeps her register qubits $(A^{(1)},...,A^{(L)})$.  Let $i$ be the position of the pulse with the detection.  Bob randomly selects $j\in\{1,...,L\}\setminus\{i\}$ and records the ordered pair $(i\rightarrow j)$.  [Success round] 
                \item If the above condition is not satisfied, Bob announces ``failure'' and Alice discards her qubits.  [Failure round]
            \end{itemize}
        \end{itemize} 
    \item Let $N$ be the number of the success rounds.  By proper indexing, Alice's qubit registers are represented by $(A_1^{(1)},..., A_1^{(L)}) ,..., (A_N^{(1)},..., A_N^{(L)})$ and Bob's records of ordered pairs are represented by $(i_{1} \rightarrow j_{1}),...,(i_{N} \rightarrow j_{N})$.
    \item Bob announces the sequence of unordered pairs $(i_{1},j_{1}), ...,(i_{{N}},j_{{N}})$.  He additionally announces the ordered pairs $(i_{1}\rightarrow j_{1}),...,(i_{N}\rightarrow j_{N})$. 
    \item According to the ordered pairs $(i_{k}\rightarrow j_{k})$ $(k \in \{1,...,N\})$, Alice applies a CNOT operation between qubits $A_k^{(i_k)}$ and $A_k^{(j_k)}$ with $A_k^{(i_k)}$ being control and $A_k^{(j_k)}$ being target.  She stores qubit $A_k^{(j_k)}$ as the $k$th sifted key qubit, which she renames as $A_k$.   {\bf She then measures qubit $A_k^{(i_k)}$ in the phase basis to obtain a binary outcome $b_k$.  She also performs phase-basis measurement on each of the $L-2$ qubits $A_k^{(l)}\ (l\in\{1,...,L\}\setminus \{i_k,j_k\}) $ to count the number $a_k \in \{0,...,L-2\}$ of the qubits with outcome $1$.  Alice records $y_k = (a_k,b_k)$.}  At the end, she has $N$ sifted key qubits $A':=A_1,...,A_N$, and the sequence $y^N:=y_1,...,y_N$.
    \item Alice chooses and announces a bit error correcting code.  
    \item Let $N_{\mathrm{fin}}:= N_{\mathrm{fin}}(N)$.  
    Alice draws a full-rank $N\times N$ binary matrix $C$ with the probability $p(C|N,N_{\mathrm{fin}})$ and announces $N \times N_{\mathrm{fin}}$ matrix $C_{PA}:=C \left(\bm{I}_\mathrm{N_\mathrm{fin}}\ \bm{O} \right)^T$.  
    She acts a unitary $ U(C) = \sum_{\bm{z}\in\{0,1\}^N}\ket{\bm{z} C }\bra{\bm{z}}_{A'} = \sum_{\bm{x}\in\{0,1\}^N}\ket{\bm{x}(C^{-1})^T{}_X}\bra{\bm{x}_X}_{A'} $ on her sifted key qubits, and performs phase basis measurement on the subsystem $A_{N_\mathrm{fin} + 1},...,A_N$ to obtain $(N - N_{\mathrm{fin}})$-bit sequence $\bm{t}$.  
    {\bf Using $y^N$ and $\bm{t}$, Alice computes ${\bm{x}^*}:=\bm{x}^{*}(N,y^N,N_{\mathrm{fin}}, \bm{t}) $} and acts a unitary $U'(\bm{x}^*) = \sum_{\bm{x}'\in \{0,1\}^{N_{\mathrm{fin}}}}\ket{(\bm{x}'\oplus\bm{x}^* \tilde{H}^{T})_X}\bra{\bm{x}'_X}_A$ on the remaining $N_\mathrm{fin}$ qubits $A:=A_{1},...,A_{N_\mathrm{fin}}$ (final key qubits), where $\tilde{H}$ is the $N_{\mathrm{fin}}\times N$ matrix $\left(\bm{I}_\mathrm{N_\mathrm{fin}}\ \bm{O} \right) C^{-1}$.  
    \item She performs bit basis measurement on the final key qubits $A$ and obtains the final key $\bm{z}_A^\mathrm{fin}$.
\end{enumerate}

We choose the parameters in Protocol 2 according to those of Protocol 1 as follows.  The constants $L$ and $N_{\mathrm{em}}$ and the function $N_{\mathrm{fin}}(N)$ are the same as those of Protocol 1.  The probability $p(C|N,N_{\mathrm{fin}})$ is chosen to satisfy
\begin{equation}
    \sum_{C:C (\bm{I}_{N_\mathrm{fin}} \ \bm{O} )^T = C_{PA}} p(C|N,N_\mathrm{fin}) = p'(C_{PA}|N,N_\mathrm{fin}).
\end{equation}

If Alice performed bit basis measurement on her register qubits of (\ref{eq:initial_state}), she obtains the random bit sequence $s^{(1)},...,s^{(L)}$ with the same probability and the $L$ optical pulses in the same state as those in Protocol 1.  In addition, all the quantum operations of Alice in Protocol 2, which are composed of permutations of the bit basis, are equivalent to the classical information processing in Protocol 1.  (Note that $U'(\bm{x}^*)$ dose not change the bit basis of the qubits.)  Furthermore, as shown in the original paper \cite{Sasaki2014}, Bob announces unordered pairs $(i_k,j_k)$ in Protocol 2 with the same probability as in Protocol 1.  Therefore, for every attack of Eve in Protocol 1, we can define a corresponding attack in Protocol 2 by letting Eve ignore the ordered pairs $(i_k\rightarrow j_k)$. 
Then, by setting the parameters as mentioned above and with the attack by Eve as defined above, we can conclude that the final state of Alice and Eve at the end of (vii) in Protocol 2 is equal to $\rho_{AE|N_\mathrm{fin}}^\mathrm{fin}$ in Protocol 1.

On the other hand, let $\rho^{\mathrm{virt}}_{AE|N_{\mathrm{fin}}}$ be the quantum state on the Alice's final key qubits $A$ and Eve's system $E$ at the end of (vi) in Protocol 2.
If $\rho^{\mathrm{virt}}_{AE|N_{\mathrm{fin}}}$ satisfies
\begin{equation}
    \sum_{N_{\mathrm{fin}}\geq 1}\mathrm{Pr}(N_{\mathrm{fin}}) \left(1 - F\left(\rho_{A|N_{\mathrm{fin}}}^{\mathrm{virt}},\ket{\bm{0}_X}\bra{\bm{0}_X}_A \right) \right) \leq \eta',
    \label{eq:upper_fidelity}
\end{equation}
where $\ket{\bm{0}_X} := \ket{0_X}^{\otimes N_{\mathrm{fin}}}$ and $\rho_{A|N_{\mathrm{fin}}}^{\mathrm{virt}} := \mathrm{Tr}_E (\rho_{AE|N_\mathrm{fin}}^\mathrm{virt})$, and Eve performs the attack as defined above, the left-hand side of (\ref{eq:secrecy}) is proved to satisfy
\begin{equation}
    \begin{aligned}
    \frac{1}{2} \sum_{N_{\mathrm{fin}}\geq 1}\mathrm{Pr}(N_{\mathrm{fin}}) \big\| \rho_{AE|N_{\mathrm{fin}}}^{\mathrm{fin}} - \rho_{AE|N_{\mathrm{fin}}}^{\mathrm{ideal}} \big\|_1 &\leq \sqrt{1 - (1 - \eta')^2} \\
    &\leq \sqrt{2\eta'},
    \end{aligned}
    \label{eq:relation_fid_trc}
\end{equation}
and thus Protocol 1 is $\sqrt{2\eta'}$-secret \cite{Koashi2009,Hayashi2012}.
 
The fidelity in the left-hand side of (\ref{eq:upper_fidelity}) is equal to the probability that Alice obtains $N_\mathrm{fin}$-bit sequence $\bm{0}:=(0 \cdots 0)$ when she measures $\rho^{\mathrm{virt}}_{AE|N_{\mathrm{fin}}}$ in the phase basis.  
We therefore consider the alternative procedure (vii)' after (vi) in Protocol 2 as follows:
\begin{enumerate}
    \item[(vii)'] 
    She performs phase basis measurement on the final key qubits $A$ and obtains the final-phase key $\bm{x}_A^\mathrm{fin}$.
\end{enumerate}
Using $\bm{x}_A^\mathrm{fin}$, the fidelity in (\ref{eq:upper_fidelity}) is given by
\begin{equation}
    F\left(\rho_{A|N_{\mathrm{fin}}}^{\mathrm{virt}},\ket{\bm{0}_X}\bra{\bm{0}_X}_A \right) = \mathrm{Pr}(\bm{x}_A^\mathrm{fin} = \bm{0}|N_\mathrm{fin}).
    \label{eq:failure_estimation}
\end{equation}
In order to evaluate the right-hand side, we introduce a third protocol which faithfully simulates the statistics of $\bm{x}_A^\mathrm{fin}$ as follows.\\

\noindent{\bf Protocol 3 (estimation protocol):}
\begin{itemize}
    \item[(i)] Alice and Bob follow the step (i) of Protocol 2 except that Alice measures the $L$ qubits in the phase basis immediately after its preparation, and obtains a bit sequence $\mathfrak{s}^{(1)},...,\mathfrak{s}^{(L)}$.  She records $m=\sum_{l=1}^{L}\mathfrak{s}^{(l)}\ (\in \{0,...,L\})$ for every round.
    In the success rounds, Alice records the sequence $(\mathfrak{s}_k^{(1)},...,\mathfrak{s}_k^{(L)})$.  Let $v_M$ be the number of rounds with $m = M$, where $\sum_M v_M = N_\mathrm{em}$.  
    \item[(ii)] By proper indexing, Alice has the bit sequences $(\mathfrak{s}^{(1)}_1,...,\mathfrak{s}^{(L)}_1),...,(\mathfrak{s}^{(1)}_N,...,\mathfrak{s}^{(L)}_N)$.  Bob has the sequence of ordered pairs, $(i_{1} \rightarrow j_{1}),...,(i_{N} \rightarrow j_{N})$.
    \item[(iii)] Bob announces the sequence of unordered pairs $(i_{1},j_{1}), ...,(i_{{N}},j_{{N}})$.  He additionally announces the ordered pairs $(i_{1}\rightarrow j_{1}),...,(i_{N}\rightarrow j_{N})$. 
    \item[(iv)] With the ordered pairs $(i_{k}\rightarrow j_{k})$ $(k \in \{1,...,N\})$, Alice computes the following variables for $k \in \{1,...,N\}$.
    \begin{eqnarray}
        x_k &:=& \mathfrak{s}^{(j_k)}_{k} \label{eq:def_x_k} \\
        m_k &:=& \sum_{l \in \{1,...,L\}}\mathfrak{s}_k^{(l)} \\
        u_k &:=& \mathfrak{s}^{(i_k)}_k \\
        a_k &:=& \sum_{l \in \{1,...,L\}\setminus\{i_k, j_k\}} \mathfrak{s}_k^{(l)} \ = m_k - u_k - x_k  \\
        b_k &:=& \mathfrak{s}^{(i_k)}_k \oplus \mathfrak{s}^{(j_k)}_{k} = u_k \oplus x_k \label{eq:def_b_k}\\
        y_k &:=& (a_k,b_k) \label{eq:def_y_k}
    \end{eqnarray}
    At the end, she has a sifted-phase key $\bm{x}_A := (x_1 \cdots x_N) (= x^N) $ and the sequence $y^N := y_1,...,y_N$ as well as the sequences $m^N := m_1,...,m_N$ and $u^N := u_1, ...,u_N$.
    \item[(vi)] She draws a full-rank $N\times N$ binary matrix $C$ with probability $p(C|N,N_\mathrm{fin})$.  She computes $H:=(\bm{O}\ \bm{I}_{N - N_\mathrm{fin}}) C^{-1}$, $\tilde{H}:= (\bm{I}_{N_\mathrm{fin}}\ \bm{O}) C^{-1}$, and $\bm{t} := \bm{x}_A H^T$.  Using $y^N$ and $\bm{t}$, she computes $\bm{x}^* :=\bm{x}^*(N,y^N,N_\mathrm{fin},\bm{t})$ and obtains the final-phase key $\bm{x}_A^\mathrm{fin} = (\bm{x}_A \oplus \bm{x}^*)\tilde{H}^{T}$. 
\end{itemize}
 
Since all the quantum operations of Alice in Protocol 2 are composed of permutations of the phase basis states, Alice's procedures of determining $\bm{x}_A^\mathrm{fin}$ in Protocol 2 with (vii)' and those in Protocol 3 are equivalent.  (We used the property of CNOT operation to derive (\ref{eq:def_x_k}) and (\ref{eq:def_b_k}).)
It is clear in Protocol 3 that the following inequality always holds:
\begin{equation}
    \mathrm{Pr}(N_\mathrm{fin}\geq 1,\bm{x}_A^\mathrm{fin} \neq \bm{0}) \leq \mathrm{Pr}(N_\mathrm{fin}\geq 1, \bm{x}_A\neq \bm{x}^{*}).
\end{equation}
With (\ref{eq:failure_estimation}), the left-hand side of the above inequality is identified as the left-hand side of (\ref{eq:upper_fidelity}).
Therefore, if we can ensure
\begin{equation}
    \mathrm{Pr}(N_\mathrm{fin}\geq 1, \bm{x}_A\neq \bm{x}^{*}) \leq \eta',
    \label{eq:misidentification}
\end{equation}
the condition (\ref{eq:upper_fidelity}) is satisfied.
The parameter $\eta'$ in (\ref{eq:misidentification}) can be regarded as an upper-bound on the probability that Alice misidentifies the sequence $\bm{x}_A$ (phase error patterns) and computes the wrong sequence $\bm{x}^*$ when given the sequence $y^N$ and the syndrome $\bm{t}$.

The bound $\eta'$ can be further related to the number of candidates of $\bm{x}_A$, given $N$ and $y^N$. 
Suppose that a family of sets $T(N,y^N)$ satisfies
\begin{equation}
    \mathrm{Pr}(N\geq 1,\bm{x}_A\notin T(N,y^N)) \leq \eta.
    \label{eq:out_of_advocate}
\end{equation} 
Suppose further that for a function $H_{PA}(N)$ which depends only on $N$,
\begin{equation}
    N H_{PA}(N) \geq {\log \bigl| T(N,y^N) \bigr|} \text{ for all } y^N 
    \label{eq:amount_PA}
\end{equation}
holds, where $\bigl| T(N,y^N) \bigr|$ is the cardinality of $T(N,y^N)$.
We assume that the selection of $H^T$ with probability $p(C|N,N_\mathrm{fin})$ in Protocol 3 is equivalent to $\text{universal}_2$ hashing, i.e.
\begin{equation}
    \begin{aligned}
    &\forall \bm{x}_1,\bm{x}_2 \in \{0,1\}^N \\
    & \ \ \  \mathrm{Pr}(\bm{x}_1 H^T = \bm{x}_2 H^T|N,N_\mathrm{fin}) \leq 2^{- (N - N_\mathrm{fin})},
    \end{aligned}
\end{equation}
which amounts to require $p'(C|N,N_\mathrm{fin})$ in Protocol 1 to be dual $\text{universal}_2$ hashing \cite{Tsurumaru2013}.
Then, by setting 
\begin{equation}
    N_{\mathrm{fin}}(N)=\max\{N(1 - H_{PA}(N) - s/N),0\},
\end{equation} 
we obtain, from the union bound,
\begin{equation}
    \mathrm{Pr}(N_\mathrm{fin}\geq 1, \bm{x}_A\neq \bm{x}^{*}) \leq \eta + 2^{-s},
    \label{eq:failure_form}
\end{equation}
because learning $\bm{t}=\bm{x}_A H^T$ eliminates all the wrong candidates in $T(N,y^N)$ except probability no more than $2^{-s}$.
Then, from (\ref{eq:upper_fidelity}), (\ref{eq:relation_fid_trc}), (\ref{eq:failure_form}), and by identifying $\eta' = \eta + 2^{-s}$, Protocol 1 is $\sqrt{2(\eta + 2^{-s})}$-secret.  

The conclusion of this subsection is as follows.  If we can define $T(N, y^N)$ which satisfies
\begin{equation}
    \mathrm{Pr}(N\geq 1,\bm{x}_A\notin T(N,y^N)) \leq \eta
    \label{eq:out_of_advocate_2}
\end{equation} 
in Protocol 3 and 
\begin{equation}
    \forall y^N, \ \  {\log \bigl| T(N,y^N) \bigr|} \leq N H_{PA}(N),
    \label{eq:amount_PA_2}
\end{equation}
for a function $H_{PA}(N)$ which depends only on $N$,
then by setting 
\begin{equation}
    N_\mathrm{fin}(N) = \max\{N(1 - H_{PA}(N) - s/N), 0\},
    \label{eq:def_N_fin}
\end{equation}
Protocol 1 can be made $\sqrt{2(\eta + 2^{-s})}$-secret.

\subsection{The origin of the improvement}

\begin{figure}[t]
    \centering
    \includegraphics[width=86mm]{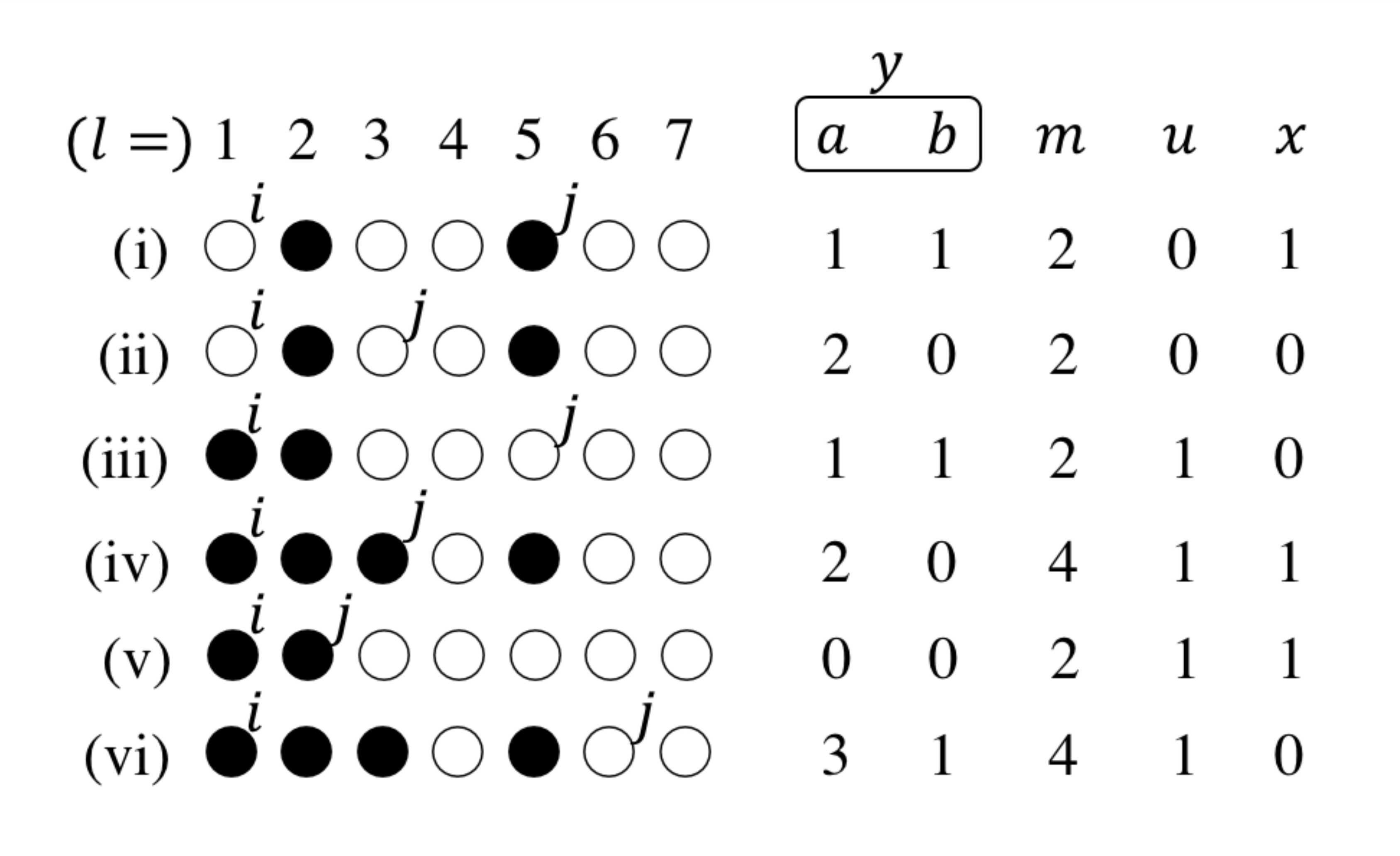}
    \caption{The illustration of how additional information $y=(a,b)$ works when $L=7$.  The white circle denotes $\mathfrak{s}^{(l)} = 0$ and the black circle denotes $\mathfrak{s}^{(l)} = 1$.  The pairs \{(i), (ii)\}, \{(iii), (v)\},  and \{(iv), (vi)\} correspond to the cases in which Eve's attacks are the same but different positions are chosen for $j$ due to the random selection by Bob.}
    \label{fig:how_tag_works}
\end{figure} 

Here we give a crude explanation of why we expect an improvement of the key rate from the introduction of additional information $y_k=(a_k,b_k)$ collected by Alice in Protocol 2.  In the asymptotic limit, the ratio $N_\mathrm{fin} / N$ is given by $N_\mathrm{fin} / N = 1 - H_{PA}$, where $H_{PA}$ is the fraction for the shortening in privacy amplification, representing an upper-bound on the amount of leaked information.  In the original security proof, for the implementation without phase randomization, it is simply given by $H_{PA}=\max h(e_\mathrm{ph})$, where $e_\mathrm{ph}$ is the average phase error probability of a sifted key qubit.  In this framework, the best strategy by Eve is to make $e_\mathrm{ph}$ as high as possible.  It is simply achieved by her measuring all the photon number parities $\mathfrak{s}^{(1)},..., \mathfrak{s}^{(L)}$, followed by choosing the index $i$ such that $\mathfrak{s}^{(i)}=0$, as illustrated in Figure \ref{fig:how_tag_works}, (i) and (ii).  Since the index $j$ is chosen randomly, phase error occurs (like (i)) with probability $m/(L-1)$ for a round, resulting in $e_\mathrm{ph} = \sum_{k}m_k / N(L - 1)$.  Hence, Eve will only have to choose $N$ rounds with higher values of $m=\sum_{l}\mathfrak{s}^{(l)}$.

The introduction of $y_k=(a_k,b_k)$ drastically changes Eve's strategy.  In this case, the asymptotic fraction will be given by a conditional entropy as $H_{PA} = \max \sum_{y} p(y)h(e_\mathrm{ph}(y))$, where $e_\mathrm{ph}(y)$ is the phase error probability conditioned on $y=(a,b)$.  As seen in Figure \ref{fig:how_tag_works}, case (i) and case (ii) have distinct values of $y$, and thus no longer contributes to $H_{PA}$.  In order to increase the conditional entropy, Eve must mix the case with the same values of $y$, such as cases (iii) and (iv).  Due to the randomness of index $j$, these inevitably lead to occurrence of other cases like (v) and (vi), and this continues.  Notice that these cases involve different values of $m=\sum_{l}\mathfrak{s}^{(l)}$.  Hence simply choosing higher values of $m$ no longer works for Eve, and she must find an appropriate balance over the values of $m$ to make the conditional entropy higher.

We emphasize here that the above constraint for Eve is quite natural once we notice that her true objective is not to increase the phase error probability but to learn the optical phase difference ${s}^{(i)}\oplus {s}^{(j)}$ between the pair of pulses.  The value of ${s}^{(i)} \oplus {s}^{(j)}$ is encoded on the relative phase of superposition states of (i) and (iii), and on that of (ii) and (iv), for example.  In this sense, the introduction of $y_k=(a_k, b_k)$ can be interpreted as providing more precise evaluation of Eve's ability to learn Alice's sifted key bits.  The reduction to Protocol 3 in the previous subsection is essentially regarded as reducing that evaluation to a problem on classical random variables possessed by Alice alone.  It is nonetheless convoluted and involves many variables and constraints, but it will be efficiently solved by introducing Lagrange multipliers in the next subsection.
       
\subsection{The estimation of the number of phase error patterns}
In this subsection, we give an explicit construction of $T(N,y^N)$, the set of likely phase error patterns.  The construction has free parameters $\bm{\nu}$ and $\bm{\xi}$ served as Lagrange multipliers, which will be defined later.  While any proper choice of the parameters makes Protocol 1 secure, the key length will depend on the choice. 

In what follows, we adopt the following notations.  For a finite set ${\cal W}$, we define ${\cal P}_{\cal W}$ as the set of all the probability mass functions on ${\cal W}$.  When a set $\Gamma$ is associated with ${\cal W}$ uniquely by a function $f_{\Gamma}:{\cal W} \rightarrow \Gamma$, we denote the distribution on $\Gamma$ induced from $P \in {\cal P}_{\cal W}$ by ${P}_{\Gamma}$, which satisfies 
\begin{equation}
    P_{\Gamma}(g) = \sum_{w \in {\cal W} : f_{\Gamma}(w)=g}P(w)
\end{equation}
for $g \in \Gamma$.  For a finite set $\Omega$, the type $\tilde{P}_{\omega^n}\in {\cal P}_{\Omega}$ for $\omega^n = (\omega_1,...,\omega_n) \in \Omega^n $ is defined by 
\begin{equation}
    \tilde{P}_{\omega^n}(\omega) = \frac{1}{n} \bigl| \{i\in \{1,...,n\} :\omega_i = \omega \} \bigr|
    \label{eq:def_type}
\end{equation} 
for $\omega \in \Omega$.

Let ${\cal M} = \{0,...,L\} ,\ {\cal U} = {\cal X} = \{0,1\}$ and ${\cal Y} = \{(a,b) | a\in\{0,...,L-2\}, \ b \in\{0,1\}\}$ be the set of all the possible values of $m_k,u_k,x_k,$ and $y_k$ in Protocol 3, respectively.
Let ${\cal W}$ be the finite set defined as follows:
\begin{equation}
    {\cal W} = \left\{(M,U,X) \in  {\cal M} \times {\cal U} \times {\cal X} : 0\leq M-U-X \leq L-2 \right\}.
\end{equation} 
Let $f_{\cal M}:{\cal W} \rightarrow {\cal M}$, $f_{\cal M \times U}:{\cal W} \rightarrow {\cal M \times U}$, and $f_{\cal X}:{\cal W} \rightarrow {\cal X}$ be the projections from the Cartesian product ${\cal M\times U \times X} $ restricted on ${\cal W}$.
Let $f_{{\cal Y}}:{\cal W} \rightarrow {\cal Y} $ be the function defined by
\begin{equation}
    f_{{\cal Y}} (M,U,X) := (M-U-X,U\oplus X).
    \label{eq:func_y}
\end{equation}
In Protocol 3, $x_k,u_k,$ and $m_k$ are related to $y_k$ by 
\begin{equation}
    y_k = f_{{\cal Y}}(m_k,u_k,x_k),
    \label{eq:relation_tag}
\end{equation} 
and hence $y^N$ is uniquely determined once sequences $x^N$, $m^N$, and $u^N$ are given.
We denote the binomial distribution with $L$ trials by $\mathfrak{b}_{L,p} \in {\cal P}_{\cal M}$, where
\begin{equation}
    \mathfrak{b}_{L,p}(M):= \binom{L}{M} p^{M} (1 - p)^{L-M}.
\end{equation}

When Eve's attack is fixed in Protocol 3, the joint probability distribution of $\bm{v}:=(v_0,...,v_L), N, x^N, m^N,$ and $u^N$, denoted by $\mathfrak{P}$, is determined.  In what follows, $\mathrm{Pr}\{ \cdot \} $ denotes the probability under $\mathfrak{P}$.  
Regardless of Eve's attacks, the following three conditions hold for $\mathfrak{P}$.
\begin{enumerate}
    \item[1.]
    The variable $\bm{v}=(v_0,...,v_L)$ obeys multinomial distribution
    \begin{equation}
        \mathrm{Pr}\{\bm{v}\} =  \frac{N_{\mathrm{em}}!}{\prod_{M=0}^{L}v_M !}\prod_{M=0}^{L}\left(\mathfrak{b}_{L,p_{\mathrm{odd}}}(M)\right)^{v_M}
        \label{eq:source_statistics}
    \end{equation} 
    with $p_\mathrm{odd}$ satisfying
    \begin{equation}
        p_\mathrm{odd} \leq p_\mathrm{src}.
        \label{eq:upper_p_odd}
    \end{equation}
    This property can be confirmed if we rewrite the initial state of Protocol 3, given by (\ref{eq:initial_state}), as
    \begin{equation}
        \begin{aligned}
        & 2^{- L/2}\bigotimes_{l=1}^{L} \sum_{s^{(l)} = 0, 1}\ket{s^{(l)}}_{A^{(l)}} \mathrm{e}^{\pi \mathrm{i} s^{(l)} \hat{n}_{l}}\ket{\Psi_{\sigma}}_{l R^{(l)}} \\
        =& \bigotimes_{l=1}^{L}\left( \ket{0_X^{(l)}}_{A^{(l)}} \frac{1 + (-1)^{\hat{n}_l}}{2}\ket{\Psi_{\sigma}}_{l R^{(l)}} \right. \\
        &\left. \hspace{30mm} + \ket{1_X^{(l)}}_{A^{(l)}} \frac{1 - (-1)^{\hat{n}_l}}{2}\ket{\Psi_{\sigma}}_{l R^{(l)}} \right) \\
        =& \bigotimes_{l=1}^{L}\left( \ket{0_X^{(l)}}_{A^{(l)}} \Pi_{l}^\mathrm{even}\ket{\Psi_{\sigma}}_{l R^{(l)}}  + \ket{1_X^{(l)}}_{A^{(l)}} \Pi_{l}^\mathrm{odd}\ket{\Psi_{\sigma}}_{l R^{(l)}} \right), 
        \end{aligned}
        \label{eq:initial_phase_basis}
    \end{equation}
    where $\Pi_{l}^\mathrm{even(odd)}$ is the projection operator onto the even (odd) photon number states.
    The probability of obtaining $\mathfrak{s}^{(l)}=1$ when measuring the $l$-th qubit of (\ref{eq:initial_phase_basis}) in phase basis, denoted by $p_{\mathrm{odd}}$, is given by
    \begin{eqnarray}
        p_{\mathrm{odd}} &:=& \mathrm{Tr}  \left( \Pi_{l}^\mathrm{odd}\ket{\Psi_{\sigma}}\bra{\Psi_{\sigma} }_{l R^{(l)}} \Pi_{l}^\mathrm{odd} \right) \nonumber \\
        &=& \mathrm{Tr} \left( \Pi_{l}^\mathrm{odd} \sigma \right).\label{eq:p_odd}
    \end{eqnarray}
    Hence the number $M = \sum_{l} \mathfrak{s^{(l)}}$ follows the probability $\mathfrak{b}_{L,p_\mathrm{odd}}(M)$.
    Since $p_\mathrm{odd}$ is equal to the probability of emitting odd number of photons in a pulse, (\ref{eq:upper_p_odd}) holds by definition. 
    \item[2.]
    For the type $\tilde{P}_{m^N} $ of the random variable $m^N$,
    \begin{equation}
        {}^{\forall}M \in {\cal M},\ \mathrm{Pr}\left\{ \left. N \tilde{P}_{m^N}(M) \leq {v_M} \right| N\geq 1 \right\} = 1 
        \label{eq:n_odd_photon}
    \end{equation}
    holds, which is obvious from the definition of the type. 
    \item[3.]
    Since Bob randomly chooses $j_k$ out of $\{1,...,L\} \setminus \{ i_k \} $ in each success round, the probability of obtaining $x_k=1$ in the $k$th success round given $m_k$ and $u_k$ is $(m_k - u_k)/(L-1)$. Therefore,
    \begin{equation}
        \begin{aligned}
        &\mathrm{Pr}\left\{ \left. x^N \right| N, m^N, u^N \right\} \\
        & \hspace{2cm}  = \prod_{k=1}^{N} [c(m_k, u_k)]^{x_k}[1 - c(m_k, u_k)]^{1 - x_k}
        \end{aligned}
        \label{eq:randomness_Bob}
    \end{equation}
    holds for $N\geq 1$, where 
    \begin{equation}
        c(M,U):=\frac{M-U}{L-1}.
    \end{equation}
\end{enumerate}
  
Let $ \bm{\nu} = \{\nu_0,...,\nu_L\} $ be the set of real non-negative constants which satisfy $\nu_M = 0$ for all $M < Lp_\mathrm{src}$.  Let ${\cal P}^{N,\bm{\nu},\delta_1} \subseteq {\cal P}_{{\cal W}}$ be the set of the probability mass functions defined by
\begin{equation}
    \begin{aligned}
    {\cal P}^{N,\bm{\nu},\delta_1} :=& \left\{P\in {\cal P}_{{\cal W}} : \mathbb{E}_{M\sim P_{{\cal M}}}[\nu_M]\right. \\
    &\hspace{2cm} \left. \leq \frac{N_{\mathrm{em}}}{N} \left( \mathbb{E}_{M \sim \mathfrak{b}_{L,p_\mathrm{src}}}[\nu_M] + \delta_1 \right) \right\}.
    \end{aligned}
    \label{eq:source_cons}
\end{equation}
From the conditions 1 and 2 of $\mathfrak{P}$, the type $\tilde{P}_{m^N,u^N,x^N} \in {\cal P}_{{\cal W}} $ in Protocol 3 belongs to ${\cal P}^{N,\bm{\nu},\delta_1}$ with a high probability.  More precisely, we have the following proposition with its proof given in Appendix A.

\begin{prop}
    Let $\bm{\nu} = \{\nu_0,...,\nu_L \} $ be the set of non-negative constants which satisfy $\nu_M=0$ for all $M < Lp_\mathrm{src}$. 
    Suppose that $\eta_1$ and $\delta_1$ satisfy
    \begin{equation}
        \max_{ Q\in{\cal Q}} 2^{-N_{\mathrm{em}} D(Q \| \mathfrak{b}_{L,p_\mathrm{src}}) } \leq \eta_1,
        \label{eq:kl_div}
    \end{equation}
    where the convex set ${\cal Q}$ is defined as
    \begin{equation}
        {\cal Q} = \left\{Q \in {\cal P}_{{\cal M}}: \mathbb{E}_{M\sim Q}[\nu_M] \geq \mathbb{E}_{M\sim \mathfrak{b}_{L,p_\mathrm{src}}}[\nu_M] + \delta_1 \right\}.
        \label{eq:def_of_Q}
    \end{equation}
    When the random variables $(\bm{v},N,m^N,u^N,x^N)$ satisfy (\ref{eq:source_statistics}) and (\ref{eq:n_odd_photon}), the following inequality holds:
    \begin{equation}
        \mathrm{Pr}\{N \geq 1, \tilde{P}_{m^N,u^N,x^N} \notin {\cal P}^{N,\bm{\nu},\delta_1}\} \leq \eta_1.
        \label{eq:out_of_kl}
    \end{equation}    
\end{prop} 

Let $\bm{\xi} = \{ \xi_{M,U}: (M,U) \in {\cal M}\times{\cal U},\ 0\leq c(M,U) \leq 1 \} $ be the set of real constants satisfying $\bigl|\xi_{M,U}\bigr| \leq 1$.  Let ${\cal P}^{\bm{\xi}, \delta_2} \subseteq {\cal P}_{{\cal W}}$ be the set of the probability mass functions defined as
\begin{widetext}
\begin{equation}
    \begin{aligned}
    {\cal P}^{\bm{\xi}, \delta_2} :=& \left\{P\in{\cal P}_{{\cal W}} : \mathbb{E}_{(M,U,X) \sim P} \left[ \left(X-c(M,U) \right) \xi_{M,U} \right] \right. \\ 
    & \hspace{3cm} \left. \leq \frac{\delta_2}{3} + \left[\left(\frac{\delta_2}{3}\right)^2 + 2\delta_2\mathbb{E}_{(M,U)\sim P_{\cal M\times U}}\left[c(M,U)\left(1 - c(M,U)\right)\xi_{M,U}^2\right] \right]^{\frac{1}{2}} \right\}.
    \label{eq:random_choice}
    \end{aligned}
\end{equation}
\end{widetext}
Since the right-hand side of the inequality is a concave function with respect to $P$, ${\cal P}^{\bm{\xi}, \delta_2}$ is a convex subset of ${\cal P}_{{\cal W}}$.
From the condition 3 of $\mathfrak{P}$, the type $\tilde{P}_{m^N,u^N,x^N} \in {\cal P}_{{\cal W}} $ in Protocol 3 belongs to ${\cal P}^{\bm{\xi}, \delta_2(N)}$ with a high probability.  More precisely, we have the following proposition with its proof given in Appendix B.
\begin{prop}
    Let $\bm{\xi} = \{ \xi_{M,U}: (M,U) \in {\cal M}\times{\cal U},\ 0\leq c(M,U) \leq 1\} $ be the set of real constants which satisfy $\bigl|\xi_{M,U}\bigr| \leq 1 $.
    Suppose that $\eta_2$ and $\{\delta_2(N)\}_{N=1,2,...}$ satisfy
    \begin{equation}
        \exp\left[-N \delta_2(N)\right] \leq \eta_2.
        \label{eq:failure_bernstein}
    \end{equation}
    When the random variables $(N,m^N,u^N,x^N)$ satisfy the condition (\ref{eq:randomness_Bob}), the following inequality holds:
    \begin{equation}
        \mathrm{Pr}\left\{N\geq 1, \tilde{P}_{m^N,u^N,x^N} \notin {\cal P}^{\bm{\xi}, \delta_2(N)}\right\} \leq \eta_2.
        \label{eq:bernstein_cons}
    \end{equation} 
\end{prop}

We define the following convex set of probability mass functions over ${\cal W}$,
\begin{equation}
    {\cal E} := \left\{ P :P\in {\cal P}^{N,\bm{\nu},\delta_1} \cap {\cal P}^{\bm{\xi}, \delta_2(N)}\right\},
    \label{eq:convex_region}
\end{equation}
which satisfies
\begin{equation}
    \mathrm{Pr}\{N\geq 1, \tilde{P}_{m^N,u^N,x^N}\notin {\cal E}\} \leq \eta_1 + \eta_2,
    \label{eq:out_of_E}
\end{equation}
if $(\eta_1, \delta_1)$ and $(\eta_2,\{\delta_2(N)\}_{N=1,...})$ satisfy (\ref{eq:kl_div}) and (\ref{eq:failure_bernstein}), respectively (union bound).
With ${\cal E}\subseteq {\cal P}_{{\cal W}}$, we define the set of likely phase error patterns $T(N,y^N)$ as follows:
\begin{equation}
    T(N,y^N) := \{ x^N \in {\cal X}^N : {}^\exists P \in {\cal E}, P_{{\cal X}\times{\cal Y}} = \tilde{P}_{x^N,y^N} \}.
    \label{eq:error_pattern}
\end{equation}
If $\tilde{P}_{m^N,u^N,x^N}\in {\cal E}$, by setting $P=\tilde{P}_{m^N,u^N,x^N}$, we have $P_{{\cal X}\times{\cal Y}} = \tilde{P}_{x^N,y^N}$, and thus $x^N\in T(N,y^N)$.
Therefore, from (\ref{eq:out_of_E}), we also have
\begin{equation}
    \mathrm{Pr}\{N \geq 1, x^N \notin T(N,y^N) \} \leq \eta_1 + \eta_2.
    \label{eq:out_of_advocate_3}
\end{equation} 
Here, the upper-bound of $\bigl| T(N,y^N) \bigr|$ is obtained by using the following lemma.

\begin{lemma}[The upper bound on the number of distinct patterns compatible to a joint probability distribution]  
    Let ${\cal W}$ be a finite set, and ${\cal E}$ be a closed convex subset of ${\cal P}_{\cal W}$.  Let ${\cal X}$ and ${\cal Y}$ be sets associated with ${\cal W}$ by functions $f_{\cal X}:{\cal W \rightarrow X}$ and $f_{\cal Y}:{\cal W \rightarrow Y}$.  For $y^N\in {{\cal Y} }^N$, define the set
    \begin{equation}
        T(y^N):= \{ x^N\in  {\cal X}^N : {}^{\exists} P \in {\cal E}, P_{\cal X \times Y} = \tilde{P}_{x^N,y^N} \}.
    \end{equation}
    Then the cardinality of the set $T(y^N)$ satisfies
    \begin{equation}
        \bigl|T(y^N)\bigr| \leq \max_{P\in {\cal E}:P_{\cal Y} = \tilde{P}_{y^N}}2^{N H(X|Y)_{P_{\cal X \times Y}}}.
        \label{eq:upper_estimation}
    \end{equation}
\end{lemma}
\noindent Although we have assumed specific choices of ${\cal W},f_{\cal X},f_{\cal Y},$ and ${\cal E}$, we can generally prove Lemma 1 without such specification, as shown in Appendix C. 
Since what we need is a bound on $\bigl|T(N,y^N)\bigr|$ independent of $y^N$ as in (\ref{eq:amount_PA_2}), we use Lemma 1 with $T(y^N)=T(N,y^N)$ and take the maximum with all the possible sequence $y^N$ as follows:
\begin{eqnarray}
        &&\max_{y^N} {\log \bigl|T(N,y^N)\bigr|} \nonumber \\
        &\leq &  \max_{y^N}\max_{P\in {\cal E}: P_{{\cal Y}}=\tilde{P}_{y^N}} N {H(X|Y)_{P_{{\cal X}\times{\cal Y}}}} \nonumber \\
        &\leq & \max_{P \in {\cal E}}\ N {H(X|Y)_{P_{{\cal X}\times{\cal Y}}}}.
        \label{eq:bound_pattern}
\end{eqnarray}
Combining Proposition 1 and 2, (\ref{eq:convex_region}), (\ref{eq:error_pattern}), (\ref{eq:out_of_advocate_3}), and (\ref{eq:bound_pattern}), we arrive at the following theorem.
        
\begin{theorem}[The main result]
    Let $\bm{\nu} = \{\nu_M:M\in{{\cal M}}\} $ be the set of non-negative constants which satisfy $\nu_M=0 \text{ for all }M < Lp_\mathrm{src}$. Let $\bm{\xi} = \{ \xi_{M,U} :  (M,U)\in{\cal M}\times{\cal U},\ 0\leq c(M,U) \leq 1\}$ be the set of real constants which satisfy $\bigl|\xi_{M,U}\bigr| \leq 1$.  Let $\eta_1$ and $\delta_1$ be non-negative numbers which satisfy
    \begin{equation}
        \max_{ Q\in {\cal P}_{\cal M}:\mathbb{E}_{M\sim Q}[\nu_M] \geq \mathbb{E}_{M\sim \mathfrak{b}}[\nu_M] + \delta_1 } 2^{ -N_{\mathrm{em}} D(Q \| \mathfrak{b}) } \leq \eta_1,
    \end{equation}
    where $\mathfrak{b} := \mathfrak{b}_{L,p_\mathrm{src}}$.  Let $\eta_2$ and $\{\delta_2(N)\}_{N=1,...}$ be non-negative numbers which satisfy
    \begin{equation}
        \exp\left[-N \delta_2(N)\right] \leq \eta_2.
    \end{equation}
    Let $H_{PA}(N)$ be a function of $N$ which satisfies
    \begin{equation}
        \max_{P \in {\cal E}}\ {H(X|Y)_{P_{{\cal X}\times{\cal Y}}}} \leq H_{PA}(N),
    \end{equation}
    where ${\cal E}$ is given in (\ref{eq:convex_region}).  Then, if the three conditions (\ref{eq:source_statistics}), (\ref{eq:n_odd_photon}), and (\ref{eq:randomness_Bob}) are satisfied, there exists $T(N,y^N)$ which satisfies
    \begin{equation}
        \mathrm{Pr}\{N \geq 1, x^N \notin T(N,y^N) \} \leq \eta_1 + \eta_2
    \end{equation} 
    in Protocol 3, and 
    \begin{equation}
        \forall y^N, \ \ {\log \bigl| T(N,y^N) \bigr|} \leq N H_{PA}(N).
    \end{equation}
\end{theorem}
Combining this and the conclusion of the section \ref{sec:reduction}, we conclude that Protocol 1 can be made $\sqrt{2(\eta_1 + \eta_2 + 2^{-s})}$-secret by setting $N_\mathrm{fin}(N)$ as in (\ref{eq:def_N_fin}).

\section{Numerical simulations}

We numerically simulate the net key gain per pulse $(N_{\mathrm{fin}} - N_{EC})/{N_{\mathrm{em}}L} = N(1 - H_{PA}(N) - s/N - f_{EC}h(e_{\mathrm{bit}})) /{N_{\mathrm{em}}L}$ of the RRDPS protocol using Theorem 1.
We set 
\begin{equation}
    NH_{PA}(N) = \lceil \max_{P \in {\cal E}}\ N{H(X|Y)_{P_{{\cal X}\times{\cal Y}}}} \rceil,
    \label{eq:PA_func}
\end{equation}
where $\lceil \cdot \rceil$ denotes the ceiling function.
Since the conditional entropy function is concave with respect to the joint probability distribution $P$, what we need is to solve the following constrained convex optimization problem:
\begin{widetext}
\begin{equation}
    \begin{aligned}
        & \underset{P \in {\cal P}_{{\cal W}}}{\text{maximize}}
        &&  H(X|Y)_{P_{{\cal X}\times {\cal Y}}}, \\
        & \text{subject to}
        && \mathbb{E}_{M \sim P_{{\cal M}}}[\nu_M]  \leq \frac{N_{\mathrm{em}}}{N} \left( \mathbb{E}_{M\sim \mathfrak{b}_{L,p_\mathrm{src}}}[\nu_M] + \delta_1 \right), \\
        & && \mathbb{E}_{(M,U,X) \sim P} \left[ \left(X-c(M,U) \right) \xi_{M,U} \right] \\
        & && \hspace{20mm} \leq \frac{\delta_2(N)}{3} + \left[\left(\frac{\delta_2(N)}{3}\right)^2 + 2\delta_2(N)\mathbb{E}_{(M,U)\sim P_{\cal M\times U}}\left[c(M,U)\left(1 - c(M,U)\right)\xi_{M,U}^2\right] \right]^{\frac{1}{2}},
    \end{aligned}
    \label{eq:convex_optimization}
\end{equation}
with a proper choice of the constants $\bm{\nu}$ and $\bm{\xi}$.
\end{widetext}

First, we consider the asymptotic limit $N_\mathrm{em}, N\rightarrow \infty$ while the block detection rate $R:=N/N_{\mathrm{em}}$ remains constant.
In this case, we can neglect $\delta_1$ and $\delta_2(N)$, and the optimization problem (\ref{eq:convex_optimization}) is reduced to the following simple form:
\begin{equation}
    \begin{aligned}
        & \underset{P \in {\cal P}_{{\cal W}}}{\text{maximize}}
        &&  H(X|Y)_{P_{{\cal X}\times {\cal Y}}}, \\
        & \text{subject to}
        && \mathbb{E}_{M \sim P_{{\cal M}}}[\nu_M]  \leq \frac{\mathbb{E}_{M\sim \mathfrak{b}_{L,p_\mathrm{src}}}[\nu_M]}{R} , \\
        & && \mathbb{E}_{(M,U,X) \sim P} \left[ \left(X-c(M,U) \right) \xi_{M,U} \right] = 0.
    \end{aligned}
    \label{eq:optimization_asympt}
\end{equation}
Here the equality of the second constraint comes from the fact that $\xi_{M,U}$ can be both positive and negative.
Finding the best bound on $H(X|Y)_{P_{{\cal X}\times {\cal Y}}}$ by adjusting $\bm{\nu}$ and $\bm{\xi}$ is equivalent to solving the following convex optimization problem with the affine constraints:
\begin{equation}
    \begin{aligned}
        & \underset{P \in {\cal P}_{{\cal W}}}{\text{maximize}}
        &&  H(X|Y)_{P_{{\cal X}\times {\cal Y}}}, \\
        & \text{subject to}
        && P_{\cal M}(M)  \leq \frac{\mathfrak{b}_{L,p_\mathrm{src}}(M)}{R}, \\
        & && \hspace{12mm} \forall M\in {\cal M},\ M\geq Lp_\mathrm{src}, \\
        & && \\
        & && P(M,U,X=1) - c(M,U)P_{\cal M\times U}(M,U) = 0, \\ 
        & && \hspace{12mm} \forall (M,U) \in {\cal M \times U},\ 0 \leq c(M,U) \leq 1.
    \end{aligned}
    \label{eq:optimization_asympt2}
\end{equation}
Since the problem is convex, if we can find $P^*(M,U,X)>0, \bm{\nu}^*, \bm{\xi}^*,$ and $\lambda^*$ that satisfy the following Karush-Kuhn-Tucker (KKT) condition, the maximum of $H(X|Y)_{P_{{\cal X}\times {\cal Y}}}$ in the problem (\ref{eq:optimization_asympt2}) is achieved at $P=P^*$.
\begin{widetext}
\begin{equation}
    \left\{
    \begin{aligned}
        &\nabla_P \left[ H(X|Y)_{P_{{\cal X}\times{\cal Y}}} -\lambda^{*} \sum_{(M,U,X) \in {\cal W}}P(M,U,X) - \sum_{M\in{\cal M}}\nu_M^* P_{{\cal M}}(M) \right.\\
        & \hspace{54mm} \left. - \sum_{\substack{(M,U)\in{\cal M}\times{\cal U} \\ 0\leq c(M,U)\leq 1}} \xi_{M,U}^{*}\left[  P(M,U,X=1) - c(M,U)P_{{\cal M}\times{\cal U}}(M,U) \right] \right]_{P=P^*} = \bm{0},  \\
        &\begin{aligned}
        \nu_M^* &= 0, && \forall M\in{\cal M},\ M<Lp_\mathrm{src}, \\
        \nu_M^* \geq 0, \ \ \ \frac{ \mathfrak{b}_{L,p_\mathrm{src}}(M)}{R} \geq P^{*}_{\cal M}(M), \ \ \ \nu_M^* \left(\frac{\mathfrak{b}_{L,p_\mathrm{src}}(M)}{R} - P^{*}_{\cal M}(M) \right) &= 0, && \forall M\in{\cal M},\ M\geq Lp_\mathrm{src}, \\
        P^{*}(M,U,X=1) - c(M,U)P^{*}_{\cal M\times U}(M,U) &= 0, && \forall (M,U)\in{\cal M}\times{\cal U},\ 0\leq c(M,U)\leq 1, \\
        \sum_{(M,U,X) \in {\cal W}}P^{*}(M,U,X) &= 1. &&
        \end{aligned}
    \end{aligned}
    \right.
    \label{eq:kkt_condition}
\end{equation}
\end{widetext}
The asymptotic limit $H_{PA}$ of the amount of privacy amplification $H_{PA}(N)$ is then given by
\begin{equation}
    H_{PA} = {H(X|Y)_{P^{*}_{{\cal X}\times{\cal Y}}}}.
\end{equation}

\begin{figure}[t]
    \centering
    \includegraphics[width=86mm]{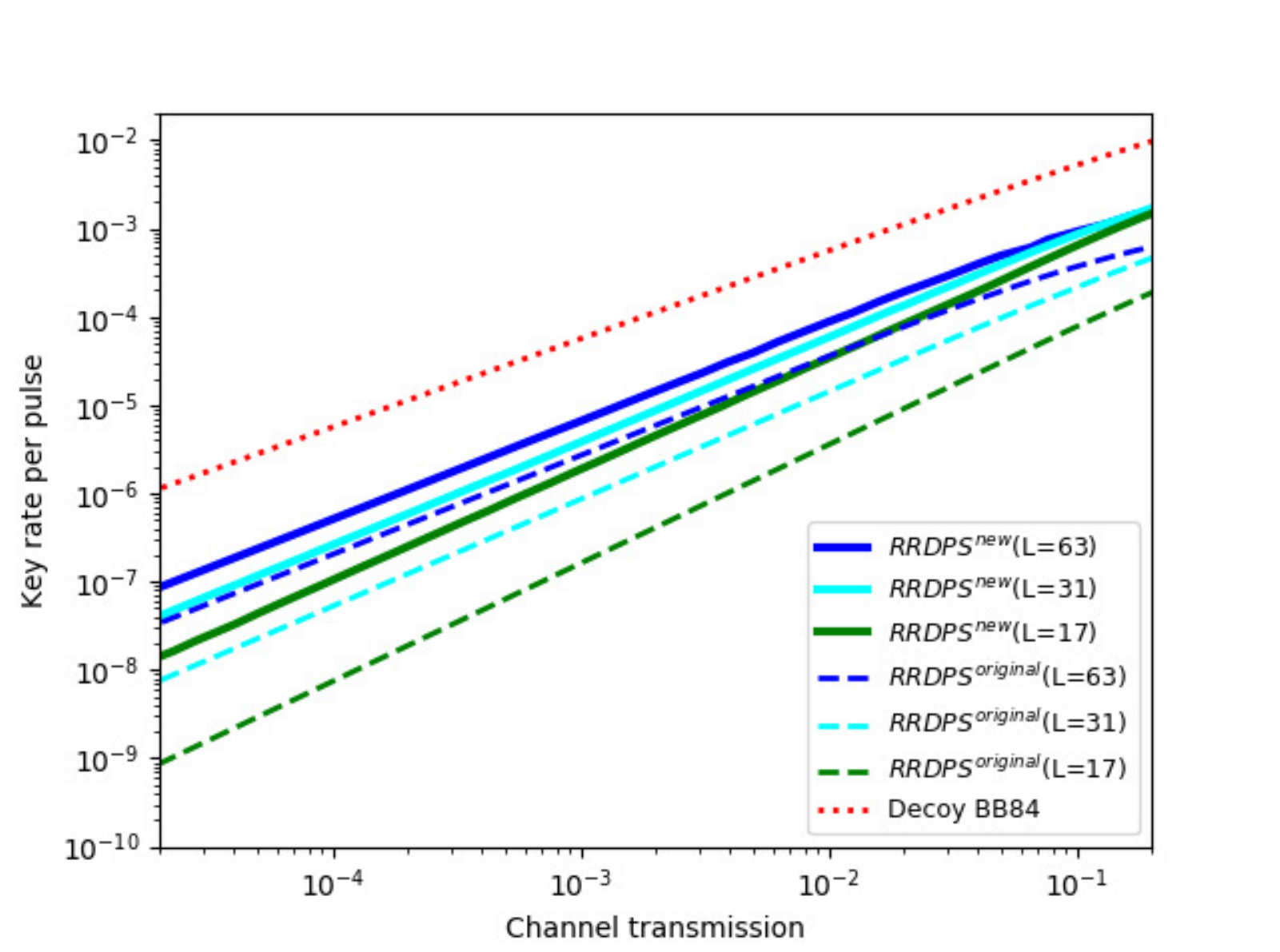}
    \caption{The asymptotic key rates of the RRDPS protocol by our new analysis ($\text{RRDPS}^{\text{new}}$, solid lines) and by the original analysis ($\text{RRDPS}^{\text{original}}$, broken lines). The mean photon number $\mu$ of the light source is optimized for each transmission rate. The bit error rate is set to 3\%. The dotted line is the rate of the ideal decoy-state BB84 protocol with time-bin implementation, assuming the same bit error rate.}
    \label{fig:keyrate_asympt}
\end{figure}

For the numerical simulation of the key rate, we assume that the block detection rate $R$ is given by
\begin{equation}
    R = \frac{1}{2}\eta \mu  L \exp(-\eta \mu L),
\end{equation}
where $\eta$ is an overall transmission rate of the channel and $\mu$ is the mean photon number of each pulse from the source.  (This rate is equal to the probability of detecting single photon in a block with efficiency $1/2$ \cite{Sasaki2014}.)  We neglect the dark count rate.  In addition, we assume that the photon number distribution of each pulse is Poissonian with mean $\mu$.  From (\ref{eq:p_odd}), $p_{\mathrm{odd}}$ in this case is given by
\begin{equation}
    p_{\mathrm{odd}} = \mathrm{e}^{-\mu}\sum_{n=0}^{\infty} \frac{\mu^{2n+1}}{(2n+1)!} = \mathrm{e}^{-\mu} \sinh \mu .
\end{equation}
We set $p_\mathrm{src} = p_\mathrm{odd} = \mathrm{e}^{-\mu} \sinh \mu$.
The error correction efficiency $f_{EC}$ is set to $1$.
We numerically solved (\ref{eq:kkt_condition}) and always found solution.  Figure \ref{fig:keyrate_asympt} shows the key rates vs. transmission rates by our new analysis and by the original analysis with the key rate of the decoy BB84 protocol with time-bin implementation when $e_{\mathrm{bit}} = 3\% $.  One can see that our new analysis improves the key rates of the RRDPS protocol for all $L$ compared to the original one.  Moreover, we obtain an improvement of more than one order of magnitude in the key rate with relatively small $L$, which may improve the practicality of the protocol.  The improved key rates with our analysis are comparable to that obtained in \cite{Yin2018}, but our analysis does not require the optical phase randomization.

\begin{figure}[b]
    \centering
    \includegraphics[width=86mm]{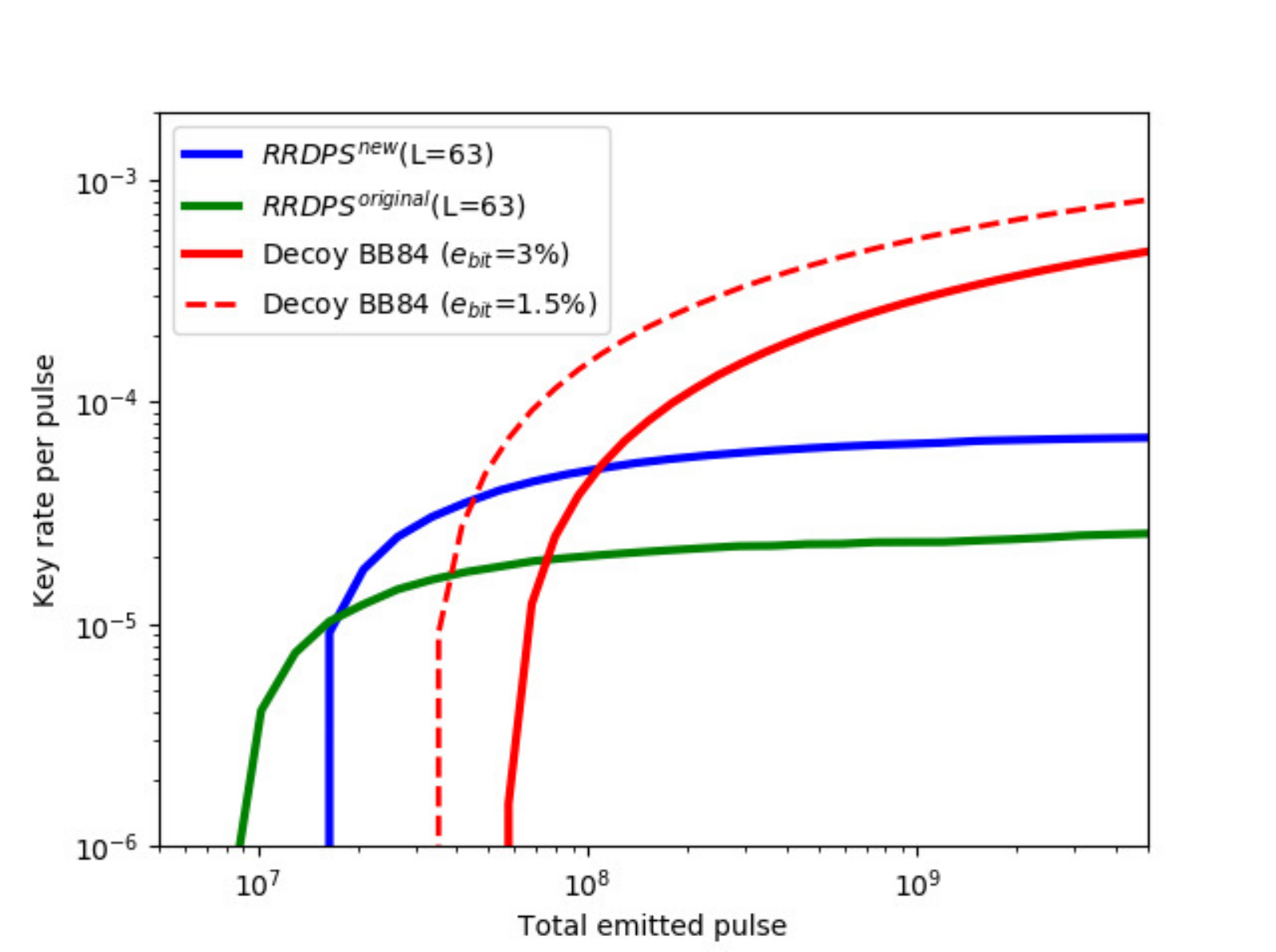}
    \caption{The key rates of the RRDPS protocol in finite-sized case with block length $L=63$.  The horizontal axis shows the total number of pulses which Alice emits.  The transmission rate is set to be $10^{-2}$, and $e_{\mathrm{bit}} = 3\% $.  The mean photon number of the light source is optimized for each number of pulses.  The red solid line is the rate of the three-state decoy BB84 protocol with time-bin implementation under the same condition \cite{Lim2014}, and the broken line is under the bit error rate 1.5\%. } 
    \label{fig:finite_size}
\end{figure}
  
Next we simulated the key rates in the finite-sized case by solving (\ref{eq:convex_optimization}) with a heuristic choice of $\bm{\nu}$ and $\bm{\xi}$.  Regardless of $N_\mathrm{em}$, we used $\{\nu_M^*:M\in{{\cal M}}\}$ and $\{ \xi_{M,U}^* : (M,U)\in{\cal M}\times{\cal U},\ 0\leq c(M,U) \leq 1 \}$, which are obtained as the solutions of (\ref{eq:kkt_condition}), to define
\begin{equation}
    \begin{aligned}
        \bm{\nu}&:=\{\nu_M^*:M\in{{\cal M}}\}, \\
        \bm{\xi}&:=\left\{ \frac{\xi_{M,U}^*}{\xi_{\max}} : (M,U)\in{\cal M}\times{\cal U},\ 0\leq c(M,U) \leq 1 \right\},
    \end{aligned}
\end{equation}
where $\xi_{\mathrm{max}}$ is defined as
\begin{equation}
    \xi_{\max}:= \max\{ \bigl|\xi_{M,U}^*\bigr| : (M,U)\in{\cal M}\times{\cal U},\ 0\leq c(M,U) \leq 1 \}.
\end{equation}
This heuristic choice of $\bm{\nu}$ and $\bm{\xi}$ becomes optimal when $N_\mathrm{em},N \rightarrow \infty$, i.e.\ in the asymptotic limit. 
We set the required correctness $\varepsilon_{\mathrm{cor}}$ and the required secrecy $\varepsilon_{\mathrm{sec}} $ to $\varepsilon_{\mathrm{cor}} = \varepsilon_{\mathrm{sec}} = 10^{-15}$.  We assumed that the bit error correction efficiency $f_{EC} = 1.2$.
Furthermore, for simplicity, we set
\begin{equation}
    \eta_1 = \eta_2 = 2^{-s} = \frac{\varepsilon_{\mathrm{sec}}^2}{6},
\end{equation}
in order to satisfy $\varepsilon_{\mathrm{sec}} = \sqrt{2(\eta_1 + \eta_2 + 2^{-s})}$.  We determined the values of $\delta_1$ and $\{\delta_2(N)\}_{N=1,...}$ by numerically solving
\begin{eqnarray}
    \eta_1 &=& \max_{ Q\in {\cal P}_{\cal M}:\mathbb{E}_{M\sim Q}[\nu_M] \geq \mathbb{E}_{M\sim \mathfrak{b}}[\nu_M] + \delta_1 } 2^{ -N_{\mathrm{em}} D(Q \| \mathfrak{b}) }, \\
    \eta_2 &=& \exp\left[-N \delta_2(N)\right],
\end{eqnarray}
where $\mathfrak{b}:= \mathfrak{b}_{L,p_\mathrm{src}}$.
When we solved the optimization problem, we used the ``minimize'' function of the scipy library in Python with the ``SLSQP'' method.
Figure \ref{fig:finite_size} shows the key rates vs. the total emitted pulses of the RRDPS protocol with our analysis and with the original analysis, and of the decoy BB84 protocol.  (The number of total emitted pulses is given by $N_{\mathrm{em}} L $ in the case of the RRDPS protocol.)  
Since the sampling cost $N_\mathrm{smp}$ is negligible only when we allow a margin for the estimation of $e_\mathrm{bit}$, we expect that the actual bit error rate should be lower than $3\%$.  For this reason, we have also plotted the rate with $e_\mathrm{bit}=1.5\%$ for the decoy BB84 protocol.
Comparison of these rates shows that the improvement of the key rates over the original proof survives up to fairly small number of total emitted pulses at which the decoy BB84 protocol fails to produce a key.

\section{Discussion}

In this paper, we proposed a refined security proof of the RRDPS protocol, which improves the key generation rate without any change in the protocol itself.  The crux of the improvement is an observation that the estimation of the phase error pattern in the virtual protocol can be aided by additional information $y^N$, which was ignored in the original security analysis.  The pair of parameters $y_k=(a_k,b_k)$ are related to the parity of the number of emitted photons in each of the $L$ pulses forming the $k$-th block (see (\ref{eq:def_x_k})-(\ref{eq:def_y_k})).  The parameter $b_k$ is associated with the pulse pair from which the sifted key bit is extracted, while the parameter $a_k$ is with the rest of $L-2$ pulses.  It is interesting that not only $b_k$ but also $a_k$ contributes to the improvement of the key rate.

The use of additional information related to the emitted numbers may look similar to the tagging technique \cite{Gottesman2004} used for the (decoy) BB84 protocols with a practical source, where the latter uses the information whether the each pulse contains multiple photons (tagged) or not (untagged).  There are, however, a couple of differences.  One is the conceptual difference arising from the timing at which the tag is defined.  In the case of the BB84 protocols, a tag is defined when the optical phase randomization is applied to the pulse before it leaves the sender.  As a result, we can easily analyze the statistical properties of the tags without regard to the Eve's attack.  In contrast, $y_k=(a_k,b_k)$ in our case should be dubbed an {\em ex post facto} tag, because it is defined only after the positions of the pair of pulses are announced by Bob.
The analysis on the statistical properties of the {\em ex post facto} tags are not straightforward and often requires special techniques to extract a property that is independent of Eve's attack \cite{Kawakami2016}.  In our case, it is solved by introducing a third protocol (Protocol 3) solely for this purpose.  From the viewpoint of implementation, the {\em ex post facto} tag has an advantage that it does not require optical phase randomization.  

Another difference is a rather technical one that becomes significant in analyzing the finite-sized case.
In contrast to the tag for the BB84 protocols which takes two values (multiple photons or not) or three values (multiple photons, single photons, or vacuum), our tag $y_k$ takes $\bigl| {\cal Y}\bigr| = 2(L-1)$ values.
As a result, the number of rounds with a specific value $y\in {\cal Y}$ of tag, $N\tilde{P}_{y^N}(y)$, is much smaller than $N$.  In addition, the constraint (\ref{eq:randomness_Bob}) essentially dictates connection between the events whose tags take different values.  In such a case, it is not wise to derive a statistical bound separately for each value of $y\in {\cal Y}$ and then to combine those bounds by using the union bound.  Instead, here we introduced Lagrange multipliers and derived an inequality for a combined property directly, as in Proposition 1 and 2.  For counting the number of phase error patterns, the bound in Lemma 1 that is independent of the size $\bigl|{\cal Y}\bigr|$ will be quite useful for mitigating finite-sized effects.
      
Although the above strategy succeeded in showing that the improvement persists up to a relatively small total number of emitted pulses, we see in Figure \ref{fig:finite_size} that the rate is eventually surpassed by the original analysis when the total number is further decreased.  We may ascribe it to either or both of the following two reasons.  One is the fact that we did not optimized the values of the Lagrange multipliers and substituted the values in the asymptotic limit instead.  The other is the use of a Bernstein's inequality in the proof of Proposition 2, which affects the key rate through the definition of ${\cal P}^{\bm{\xi},\delta_2}$ in (\ref{eq:random_choice}).  It remains to be open whether we can replace it with a tighter bound while keeping the convexity of ${\cal P}^{\bm{\xi},\delta_2}$.

We emphasize here that our focus in the present paper was on the simplest implementation of the sender's apparatus, namely, the use of the protocol without block-wise phase randomization in the original proposal, as it is.  
A natural question is that how the situation changes if we combine block-wise phase randomization with our analysis.  On one hand, the aid of block-wise phase randomization does not seem to change the key rate so much considering that the key rate with our analysis is comparable to that with \cite{Yin2018} in which the block-wise phase randomization is used.  On the other hand, ${\cal P}^{N,\bm{\nu},\delta_1}$ may be significantly narrowed by the additional photon number constraint which is the consequence of the block-wise phase randomization.  We leave it for the future research.

\begin{acknowledgements}
This work was funded in part by ImPACT Program of Council for Science, Technology and Innovation (Cabinet Office, Government of Japan), Photon Frontier Network Program (Ministry of Education, Culture, Sports, Science and Technology), CREST (Japan Science and Technology Agency), and JSPS KAKENHI Grant Number JP18K13469.
\end{acknowledgements}
        
\appendix

\section{Proof of Proposition 1}

Let $P_{\bm{v}}\in {\cal P}_{\cal M}$ be defined as $P_{\bm{v}}(M):=v_{M}/N_\mathrm{em}$.
From the condition (\ref{eq:source_statistics}) and the fact that ${\cal Q}$ is a convex set, we can apply the special case of the Sanov's theorem \cite{Csiszar1984,Sanov1957} to $P_{\bm{v}}$ as follows: 
\begin{equation}
    \mathrm{Pr}\left\{ P_{\bm{v}} \in {\cal Q} \right\}  \leq \max_{Q\in {\cal Q}}2^{-N_{\mathrm{em}}D(Q \| \mathfrak{b}_{L,p_\mathrm{odd}})}.
    \label{eq:sanov}
\end{equation}

Let $b_\mathrm{norm}$ be the constant given by
\begin{equation}
    b_\mathrm{norm} := \sum_{M: M < Lp_\mathrm{src}} \mathfrak{b}_{L,p_\mathrm{odd}}(M).
\end{equation}
Let $q \in {\cal P}_{\cal M}$ be the probability mass function which is defined as
\begin{equation}
    q(M):= \left\{
    \begin{aligned}
        &\frac{\mathfrak{b}_{L,p_\mathrm{src}}(M)}{b_\mathrm{norm}} && \forall M < Lp_\mathrm{src} \\
        &\frac{\mathfrak{b}_{L,p_\mathrm{src}}(M) - \mathfrak{b}_{L,p_\mathrm{odd}}(M)}{b_\mathrm{norm}} && \forall M \geq Lp_\mathrm{src},
    \end{aligned}
    \right.
\end{equation}
which is well-defined since $\mathfrak{b}_{L,p_\mathrm{src}}(M) \geq \mathfrak{b}_{L,p_\mathrm{odd}}(M)$ holds for all $M \geq Lp_\mathrm{src}$, and $\sum_{M\in{\cal M}}q(M) = 1$.
We define the stochastic map $S:{\cal P}_{\cal M} \rightarrow {\cal P}_{\cal M}$ as follows:
\begin{equation}
    \forall P \in {\cal P}_{\cal M},\ \ \  S(P)(M') := \sum_{M\in{\cal M}}\mathrm{Pr}(M'|M)P(M),
\end{equation}
where
\begin{equation}
    \mathrm{Pr}(M'|M) :=
    \begin{cases}
        q(M') & \forall M < Lp_\mathrm{src} \\
        \delta_{M' M} & \forall M \geq Lp_\mathrm{src}.
    \end{cases}
\end{equation}
It is easy to observe that
\begin{equation}
    \mathfrak{b}_{L,p_\mathrm{src}} = S(\mathfrak{b}_{L,p_\mathrm{odd}}).
    \label{eq:conversion}
\end{equation}
Furthermore, since (i) $S(P)(M) \geq P(M)$ and $\nu_M \geq 0$ for all $M \geq Lp_\mathrm{src}$, and (ii) $\nu_M = 0$ for all $M < Lp_\mathrm{src}$, we have
\begin{equation}
    \forall P \in {\cal P}_{\cal M},\ \ \ \mathbb{E}_{M \sim S(P)}[\nu_M] \geq \mathbb{E}_{M \sim P}[\nu_M].
    \label{eq:increasing}
\end{equation}
Therefore, from the definition of ${\cal Q}$ in (\ref{eq:def_of_Q}), we have
\begin{equation}
    \forall Q\in{\cal Q},\ \ \ S(Q)\in {\cal Q}.
    \label{eq:in_Q}
\end{equation}
Combining (\ref{eq:conversion}) and (\ref{eq:in_Q}) with the monotonicity property of the Kullback-Leibler divergence under the stochastic map \cite{Cover2012}, we have
\begin{eqnarray}
    \min_{Q\in {\cal Q}}{D(Q \| \mathfrak{b}_{L,p_\mathrm{odd}}) } &\geq& \min_{Q \in {\cal Q}} { D\left(S(Q) \| S(\mathfrak{b}_{L,p_\mathrm{odd}})\right)} \nonumber \\
    &\geq& \min_{Q \in {\cal Q}} { D(Q \| \mathfrak{b}_{L,p_\mathrm{src}})}.
    \label{eq:sanov_2}
\end{eqnarray}
From (\ref{eq:sanov}) and (\ref{eq:sanov_2}), we have
\begin{equation}
    \mathrm{Pr}\left\{ P_{\bm{v}} \in {\cal Q} \right\} \leq \max_{Q \in {\cal Q}} 2^{-N_\mathrm{em} D(Q \| \mathfrak{b}_{L,p_\mathrm{src}})}.
    \label{eq:sanov_3}
\end{equation}

On the other hand, since the random variables $(\bm{v},N,m^N)$ obey (\ref{eq:n_odd_photon}) and $\nu_M \ (M\in {\cal M})$ are non-negative, we have
\begin{equation}
    \begin{aligned}
    &\forall M \in {\cal M},\\ 
    & \hspace{5mm} \mathrm{Pr}\left\{\left.\nu_M \tilde{P}_{m^N}(M) \leq \nu_M \frac{N_\mathrm{em}}{N} P_{\bm{v}}(M)\right| N\geq 1 \right\} = 1,
    \label{eq:always_larger}
    \end{aligned}
\end{equation}
and hence
\begin{equation}
    \mathrm{Pr}\left\{ \left. \mathbb{E}_{M\sim \tilde{P}_{m^N}} [\nu_M]  \leq \frac{N_\mathrm{em}}{N}\mathbb{E}_{M\sim P_{\bm{v}}}[\nu_M]  \right| N\geq 1 \right\} = 1.
    \label{eq:sufficient_condition}
\end{equation}
Combining (\ref{eq:sufficient_condition}) with the definition of ${\cal Q}$, we have
\begin{eqnarray}
    &&\mathrm{Pr}\left\{N\geq 1, \mathbb{E}_{M\sim \tilde{P}_{m^N}} [\nu_M] \right. \nonumber\\
    &&\left. \hspace{25mm} \geq \frac{N_\mathrm{em}}{N} (\mathbb{E}_{M\sim \mathfrak{b}_{L,p_\mathrm{src}}}[\nu_M] + \delta_1)\right\} \nonumber \\
    &\leq & \mathrm{Pr}\left\{N\geq 1, \frac{N_\mathrm{em}}{N}\mathbb{E}_{M\sim P_{\bm{v}}}[\nu_M]\right. \nonumber \\
    &&\left. \hspace{25mm} \geq \frac{N_\mathrm{em}}{N} (\mathbb{E}_{M\sim \mathfrak{b}_{L,p_\mathrm{src}}}[\nu_M] + \delta_1) \right\} \nonumber \\
    &\leq & \mathrm{Pr}\{P_{\bm{v}} \in {\cal Q}\} ,
\end{eqnarray}
where the first inequality follows from (\ref{eq:sufficient_condition}).
Combining this with (\ref{eq:sanov_3}), we have
\begin{equation}
    \begin{aligned}
    &\mathrm{Pr}\left\{N\geq 1, \mathbb{E}_{M\sim \tilde{P}_{m^N}} [\nu_M] \geq \frac{N_\mathrm{em}}{N} (\mathbb{E}_{M\sim \mathfrak{b}_{L,p_\mathrm{src}}}[\nu_M] + \delta_1)\right\}\\
    & \hspace{45mm} \leq \max_{Q \in {\cal Q}} 2^{-N_\mathrm{em} D(Q \| \mathfrak{b}_{L,p_\mathrm{src}})}.
    \end{aligned}
    \label{eq:prop1_2}
\end{equation}
Then (\ref{eq:kl_div}) implies (\ref{eq:out_of_kl}).

\section{Proof of Proposition 2}

We use one of the Bernstein's inequalities \cite{Bernstein1924}, which is stated as follows.  Let $X_1,...,X_N$ be independent zero-mean random variables.  Suppose that $\bigl|X_k\bigr|\leq 1$ for all $k$.  Then, for all non-negative $t$,
\begin{equation}
    \mathrm{Pr}\left( \frac{1}{N}\sum_{k=1}^{N}X_k \geq t \right) \leq \exp\left[ - \frac{Nt^2}{\frac{2}{N}\sum_{k}\mathbb{E}[X_k^2] + \frac{2}{3} t}\right]
    \label{eq:bernstein}
\end{equation}
holds. 

For fixed values of $N(\geq 1),m^N,u^N$, the condition (\ref{eq:randomness_Bob}) determines the conditional statistics of $N$ variables $\{X_k :=\left(x_k-c(m_k,u_k)\right)\xi(m_k,u_k)\}_{k=1,...,N}$, where $\xi(M,U):=\xi_{M,U}$.  They are independent and zero-mean.  Furthermore, since $\bigl|\xi(m_k,u_k)\bigr| \leq 1$ and $0\leq c(m_k,u_k) \leq 1$, $\bigl|X_k\bigr|\leq 1$ holds for all $k$.  Thus, (\ref{eq:bernstein}) holds if we interpret $\mathrm{Pr}(\cdot)$ and $\mathbb{E}[\cdot]$ as the conditional probability and the conditional mean.
Using the definition of the type, we can rewrite the sums over index $k$ as
\begin{widetext}
\begin{eqnarray}
        \sum_{k=1}^{N}X_k
        &=& \sum_{(M,U,X)\in{\cal W}}\left(X-c(M,U)\right)\xi_{M,U} N \tilde{P}_{m^N,u^N,x^N}(M,U,X) \nonumber \\ 
        &=& N\mathbb{E}_{(M,U,X)\sim \tilde{P}_{m^N,u^N,x^N}}\left[\left(X-c(M,U)\right)\xi_{M,U}\right],
    \label{eq:prop2_1}
\end{eqnarray}
and
\begin{eqnarray}
    && \sum_{k=1}^{N}\mathbb{E}[X_k^2] \nonumber \\
    &=& \sum_{(M,U)\in{\cal M \times U}}\left\{\left[\left(0-c(M,U)\right)\xi_{M,U}\right]^2  \left(1 - c(M,U)\right) + \left[\left(1-c(M,U)\right)\xi_{M,U}\right]^2  c(M,U)\right\} N \tilde{P}_{m^N,u^N}(M,U) \nonumber \\
    &=& \sum_{(M,U)\in{\cal M \times U}} c(M,U)\left(1 - c(M,U)\right)\xi_{M,U}^2N \tilde{P}_{m^N,u^N}(M,U)\nonumber \\
    &=& N\mathbb{E}_{(M,U)\sim \tilde{P}_{m^N,u^N}}[c(M,U)(1 - c(M,U))\xi_{M,U}^2].
    \label{eq:prop2_2}
\end{eqnarray}

We choose $t$ to be 
\begin{equation}
    t = \frac{\delta_2(N)}{3} + \left[\left(\frac{\delta_2(N)}{3}\right)^2 + \frac{2\delta_2(N)}{N}\sum_{i}\mathbb{E}[X_k^2]\right]^{\frac{1}{2}},
    \label{eq:prop2_3}
\end{equation}
which satisfies
\begin{equation}
    t^2 = \delta_2(N) \left(\frac{2}{N}\sum_{i}\mathbb{E}[X_k^2] + \frac{2}{3} t\right).
\end{equation}
Substituting (\ref{eq:prop2_1}), (\ref{eq:prop2_2}), (\ref{eq:prop2_3}) to (\ref{eq:bernstein}), we obtain the following:
\begin{equation}
    \begin{aligned}
    &\mathrm{Pr}\left\{ \mathbb{E}_{(M,U,X)\sim \tilde{P}_{m^N,u^N,x^N}}\left[\left(X-c(M,U)\right)\xi_{M,U}\right] \right. \\
    & \left. \hspace{20mm} \geq \frac{\delta_2(N)}{3} + \left[\left(\frac{\delta_2(N)}{3}\right)^2 + 2\delta_2(N)\mathbb{E}_{(M,U)\sim \tilde{P}_{m^N,u^N}}\left[c(M,U)\left(1 - c(M,U)\right)\xi_{M,U}^2\right] \right]^{\frac{1}{2}} \right\} \leq \exp\left[-N\delta_2(N)\right].
    \end{aligned}
\end{equation}
Then (\ref{eq:failure_bernstein}) implies (\ref{eq:bernstein_cons}).
\end{widetext}
        
\section{Proof of Lemma 1}

For $y^N\in {{\cal Y} }^N$, define a set
\begin{equation}
    {\cal E}_{{\cal X} \times {\cal Y}} (y^N) :=\{P_{{\cal X} \times {\cal Y}}: P\in {\cal E}, P_{{\cal Y}}=\tilde{P}_{y^N} \}.
    \label{eq:projection_of_E}
\end{equation}
Since ${\cal E}$ is a closed convex set, ${\cal E}_{{\cal X} \times {\cal Y}} (y^N)$ is also a closed convex set.
Using the set, we can rewrite $T(y^N)$ as
\begin{equation}
    T(y^N):= \{ x^N\in  {\cal X}^N : \tilde{P}_{x^N, y^N} \in {\cal E}_{{\cal X} \times {\cal Y}} (y^N) \}.
\end{equation}
Consider a probability mass function $Q(x,y)$ given by
\begin{equation}
Q(x,y):=\bigl|{\cal X}\bigr|^{-1} \tilde{P}_{y^N}(y).
\end{equation}
Then we have 
\begin{equation}
\sum_{x^N\in T(y^N)} Q^N(x^N,y^N) = \bigl|{\cal X}\bigr|^{-N} \tilde{P}_{y^N}^N(y^N) \bigl|T(y^N)\bigr|
\label{eq:middle_stage}
\end{equation}
Let 
\begin{equation}
    P^* := \arg\min_{P\in {\cal E}_{{\cal X} \times {\cal Y}} (y^N)} D(P \| Q).
\end{equation}
Then, we have (Pythagorean theorem \cite{Cover2012})
\begin{equation}
    D(P \| Q)\geq D(P \| P^*) + D(P^* \| Q) \text{ for } {}^{\forall}P\in {\cal E}_{{\cal X} \times {\cal Y}} (y^N).
\end{equation}
For $(x^N, y^N)\in {\cal X}^N \times {\cal Y}^N $ with $\tilde{P}_{x^N, y^N} \in {\cal E}_{{\cal X} \times {\cal Y}}$, we have 
\begin{eqnarray}
    &&\log Q^N(x^N,y^N) - \log P^{* N}(x^N,y^N) \nonumber \\
    &=& -ND(\tilde{P}_{x^N, y^N} \| Q) + ND(\tilde{P}_{x^N, y^N} \| P^*) \nonumber  \\
    &\leq& - ND(P^* \| Q),
    \label{eq:product_prob}
\end{eqnarray}
and hence
\begin{eqnarray}
    &&\sum_{x^N\in T(y^N)} Q^N(x^N,y^N) \nonumber \\ 
    &\leq& 2^{-ND(P^* \| Q)} \sum_{x^N\in T(y^N)} P^{* N}(x^N,y^N) \nonumber \\
    &\leq& 2^{-ND(P^* \| Q)} \sum_{x^N\in {\cal X}^N}  P^{* N}(x^N,y^N) \nonumber \\
    &=& 2^{-ND(P^* \| Q)} \tilde{P}_{y^N}^N (y^N).
\end{eqnarray}
Combined with (\ref{eq:middle_stage}), we have 
\begin{equation}
 |T(y^N)| \leq 2^{-ND(P^* \| Q)+N \log  |{\cal X}|}.
 \label{eq:min_kl_div}
\end{equation}
On the other hand, for $P\in {\cal E}_{{\cal X} \times {\cal Y}} (y^N)$, 
\begin{eqnarray}
    D(P \| Q)&=&\sum_{(x,y)\in {\cal X} \times {\cal Y}} P(x,y) \log \frac{P(x,y)}{Q(x,y)} \nonumber \\
    &=&\sum_{(x,y)\in {\cal X} \times {\cal Y}} P(x,y) \log \frac{P(x|y)}{Q(x|y)} \nonumber \\
    &=& \log |{\cal X}| - H(X|Y)_P,
\end{eqnarray}
and hence
\begin{eqnarray}
    D(P^* \| Q) &=& \min_{P\in {\cal E}_{{\cal X} \times {\cal Y}} (y^N)} D(P\|Q) \nonumber \\
    &=& \log |{\cal X}| - \max_{P\in {\cal E}_{{\cal X} \times {\cal Y}} (y^N)} H(X|Y)_P.
    \label{eq:max_cond_ent}
\end{eqnarray}
Combining (\ref{eq:projection_of_E}), (\ref{eq:min_kl_div}), and (\ref{eq:max_cond_ent}) leads to (\ref{eq:upper_estimation}). 
\vspace{25mm}

\bibliography{rrdps_new}

\begin{thebibliography}{29}%
\makeatletter
\providecommand \@ifxundefined [1]{%
 \@ifx{#1\undefined}
}%
\providecommand \@ifnum [1]{%
 \ifnum #1\expandafter \@firstoftwo
 \else \expandafter \@secondoftwo
 \fi
}%
\providecommand \@ifx [1]{%
 \ifx #1\expandafter \@firstoftwo
 \else \expandafter \@secondoftwo
 \fi
}%
\providecommand \natexlab [1]{#1}%
\providecommand \enquote  [1]{``#1''}%
\providecommand \bibnamefont  [1]{#1}%
\providecommand \bibfnamefont [1]{#1}%
\providecommand \citenamefont [1]{#1}%
\providecommand \href@noop [0]{\@secondoftwo}%
\providecommand \href [0]{\begingroup \@sanitize@url \@href}%
\providecommand \@href[1]{\@@startlink{#1}\@@href}%
\providecommand \@@href[1]{\endgroup#1\@@endlink}%
\providecommand \@sanitize@url [0]{\catcode `\\12\catcode `\$12\catcode
  `\&12\catcode `\#12\catcode `\^12\catcode `\_12\catcode `\%12\relax}%
\providecommand \@@startlink[1]{}%
\providecommand \@@endlink[0]{}%
\providecommand \url  [0]{\begingroup\@sanitize@url \@url }%
\providecommand \@url [1]{\endgroup\@href {#1}{\urlprefix }}%
\providecommand \urlprefix  [0]{URL }%
\providecommand \Eprint [0]{\href }%
\providecommand \doibase [0]{https://doi.org/}%
\providecommand \selectlanguage [0]{\@gobble}%
\providecommand \bibinfo  [0]{\@secondoftwo}%
\providecommand \bibfield  [0]{\@secondoftwo}%
\providecommand \translation [1]{[#1]}%
\providecommand \BibitemOpen [0]{}%
\providecommand \bibitemStop [0]{}%
\providecommand \bibitemNoStop [0]{.\EOS\space}%
\providecommand \EOS [0]{\spacefactor3000\relax}%
\providecommand \BibitemShut  [1]{\csname bibitem#1\endcsname}%
\let\auto@bib@innerbib\@empty
\bibitem [{\citenamefont {Bennett}\ and\ \citenamefont
  {Brassard}(1984)}]{Bennett1984}%
  \BibitemOpen
  \bibfield  {author} {\bibinfo {author} {\bibfnamefont {C.~H.}\ \bibnamefont
  {Bennett}}\ and\ \bibinfo {author} {\bibfnamefont {G.}~\bibnamefont
  {Brassard}},\ }\bibfield  {title} {\bibinfo {title} {{Quantum cryptography:
  Public key distribution and coin tossing}},\ }in\ \href@noop {} {\emph
  {\bibinfo {booktitle} {Proceedings of IEEE International Conference on
  Computers, Systems, and Signal Processing}}}\ (\bibinfo {address} {India},\
  \bibinfo {year} {1984})\ p.\ \bibinfo {pages} {175}\BibitemShut {NoStop}%
\bibitem [{\citenamefont {Vallone}\ \emph {et~al.}(2015)\citenamefont
  {Vallone}, \citenamefont {Bacco}, \citenamefont {Dequal}, \citenamefont
  {Gaiarin}, \citenamefont {Luceri}, \citenamefont {Bianco},\ and\
  \citenamefont {Villoresi}}]{Vallone2015}%
  \BibitemOpen
  \bibfield  {author} {\bibinfo {author} {\bibfnamefont {G.}~\bibnamefont
  {Vallone}}, \bibinfo {author} {\bibfnamefont {D.}~\bibnamefont {Bacco}},
  \bibinfo {author} {\bibfnamefont {D.}~\bibnamefont {Dequal}}, \bibinfo
  {author} {\bibfnamefont {S.}~\bibnamefont {Gaiarin}}, \bibinfo {author}
  {\bibfnamefont {V.}~\bibnamefont {Luceri}}, \bibinfo {author} {\bibfnamefont
  {G.}~\bibnamefont {Bianco}},\ and\ \bibinfo {author} {\bibfnamefont
  {P.}~\bibnamefont {Villoresi}},\ }\bibfield  {title} {\bibinfo {title}
  {Experimental satellite quantum communications},\ }\href@noop {} {\bibfield
  {journal} {\bibinfo  {journal} {Phys. Rev. Lett.}\ }\textbf {\bibinfo
  {volume} {115}},\ \bibinfo {pages} {040502} (\bibinfo {year}
  {2015})}\BibitemShut {NoStop}%
\bibitem [{\citenamefont {Liao}\ \emph {et~al.}(2017)\citenamefont {Liao},
  \citenamefont {Cai}, \citenamefont {Liu}, \citenamefont {Zhang},
  \citenamefont {Li}, \citenamefont {Ren}, \citenamefont {Yin}, \citenamefont
  {Shen}, \citenamefont {Cao},\ and\ \citenamefont {Li}}]{Liao2017}%
  \BibitemOpen
  \bibfield  {author} {\bibinfo {author} {\bibfnamefont {S.-K.}\ \bibnamefont
  {Liao}}, \bibinfo {author} {\bibfnamefont {W.-Q.}\ \bibnamefont {Cai}},
  \bibinfo {author} {\bibfnamefont {W.-Y.}\ \bibnamefont {Liu}}, \bibinfo
  {author} {\bibfnamefont {L.}~\bibnamefont {Zhang}}, \bibinfo {author}
  {\bibfnamefont {Y.}~\bibnamefont {Li}}, \bibinfo {author} {\bibfnamefont
  {J.-G.}\ \bibnamefont {Ren}}, \bibinfo {author} {\bibfnamefont
  {J.}~\bibnamefont {Yin}}, \bibinfo {author} {\bibfnamefont {Q.}~\bibnamefont
  {Shen}}, \bibinfo {author} {\bibfnamefont {Y.}~\bibnamefont {Cao}},\ and\
  \bibinfo {author} {\bibfnamefont {Z.-P.}\ \bibnamefont {Li}},\ }\bibfield
  {title} {\bibinfo {title} {Satellite-to-ground quantum key distribution},\
  }\href@noop {} {\bibfield  {journal} {\bibinfo  {journal} {Nature}\ }\textbf
  {\bibinfo {volume} {549}},\ \bibinfo {pages} {43} (\bibinfo {year}
  {2017})}\BibitemShut {NoStop}%
\bibitem [{\citenamefont {Sasaki}\ \emph {et~al.}(2014)\citenamefont {Sasaki},
  \citenamefont {Yamamoto},\ and\ \citenamefont {Koashi}}]{Sasaki2014}%
  \BibitemOpen
  \bibfield  {author} {\bibinfo {author} {\bibfnamefont {T.}~\bibnamefont
  {Sasaki}}, \bibinfo {author} {\bibfnamefont {Y.}~\bibnamefont {Yamamoto}},\
  and\ \bibinfo {author} {\bibfnamefont {M.}~\bibnamefont {Koashi}},\
  }\bibfield  {title} {\bibinfo {title} {Practical quantum key distribution
  protocol without monitoring signal disturbance},\ }\href@noop {} {\bibfield
  {journal} {\bibinfo  {journal} {Nature}\ }\textbf {\bibinfo {volume} {509}},\
  \bibinfo {pages} {475} (\bibinfo {year} {2014})}\BibitemShut {NoStop}%
\bibitem [{\citenamefont {Takesue}\ \emph {et~al.}(2015)\citenamefont
  {Takesue}, \citenamefont {Sasaki}, \citenamefont {Tamaki},\ and\
  \citenamefont {Koashi}}]{Takesue2015}%
  \BibitemOpen
  \bibfield  {author} {\bibinfo {author} {\bibfnamefont {H.}~\bibnamefont
  {Takesue}}, \bibinfo {author} {\bibfnamefont {T.}~\bibnamefont {Sasaki}},
  \bibinfo {author} {\bibfnamefont {K.}~\bibnamefont {Tamaki}},\ and\ \bibinfo
  {author} {\bibfnamefont {M.}~\bibnamefont {Koashi}},\ }\bibfield  {title}
  {\bibinfo {title} {Experimental quantum key distribution without monitoring
  signal disturbance},\ }\href@noop {} {\bibfield  {journal} {\bibinfo
  {journal} {Nature Photonics}\ }\textbf {\bibinfo {volume} {9}},\ \bibinfo
  {pages} {827} (\bibinfo {year} {2015})}\BibitemShut {NoStop}%
\bibitem [{\citenamefont {Wang}\ \emph {et~al.}(2015)\citenamefont {Wang},
  \citenamefont {Yin}, \citenamefont {Chen}, \citenamefont {He}, \citenamefont
  {Song}, \citenamefont {Li}, \citenamefont {Zhang}, \citenamefont {Zhou},
  \citenamefont {Guo},\ and\ \citenamefont {Han}}]{Wang2015}%
  \BibitemOpen
  \bibfield  {author} {\bibinfo {author} {\bibfnamefont {S.}~\bibnamefont
  {Wang}}, \bibinfo {author} {\bibfnamefont {Z.-Q.}\ \bibnamefont {Yin}},
  \bibinfo {author} {\bibfnamefont {W.}~\bibnamefont {Chen}}, \bibinfo {author}
  {\bibfnamefont {D.-Y.}\ \bibnamefont {He}}, \bibinfo {author} {\bibfnamefont
  {X.-T.}\ \bibnamefont {Song}}, \bibinfo {author} {\bibfnamefont {H.-W.}\
  \bibnamefont {Li}}, \bibinfo {author} {\bibfnamefont {L.-J.}\ \bibnamefont
  {Zhang}}, \bibinfo {author} {\bibfnamefont {Z.}~\bibnamefont {Zhou}},
  \bibinfo {author} {\bibfnamefont {G.-C.}\ \bibnamefont {Guo}},\ and\ \bibinfo
  {author} {\bibfnamefont {Z.-F.}\ \bibnamefont {Han}},\ }\bibfield  {title}
  {\bibinfo {title} {Experimental demonstration of a quantum key distribution
  without signal disturbance monitoring},\ }\href@noop {} {\bibfield  {journal}
  {\bibinfo  {journal} {Nature Photonics}\ }\textbf {\bibinfo {volume} {9}},\
  \bibinfo {pages} {832} (\bibinfo {year} {2015})}\BibitemShut {NoStop}%
\bibitem [{\citenamefont {Guan}\ \emph {et~al.}(2015)\citenamefont {Guan},
  \citenamefont {Cao}, \citenamefont {Liu}, \citenamefont {Shen-Tu},
  \citenamefont {Pelc}, \citenamefont {Fejer}, \citenamefont {Peng},
  \citenamefont {Ma}, \citenamefont {Zhang},\ and\ \citenamefont
  {Pan}}]{Guan2015}%
  \BibitemOpen
  \bibfield  {author} {\bibinfo {author} {\bibfnamefont {J.-Y.}\ \bibnamefont
  {Guan}}, \bibinfo {author} {\bibfnamefont {Z.}~\bibnamefont {Cao}}, \bibinfo
  {author} {\bibfnamefont {Y.}~\bibnamefont {Liu}}, \bibinfo {author}
  {\bibfnamefont {G.-L.}\ \bibnamefont {Shen-Tu}}, \bibinfo {author}
  {\bibfnamefont {J.~S.}\ \bibnamefont {Pelc}}, \bibinfo {author}
  {\bibfnamefont {M.~M.}\ \bibnamefont {Fejer}}, \bibinfo {author}
  {\bibfnamefont {C.-Z.}\ \bibnamefont {Peng}}, \bibinfo {author}
  {\bibfnamefont {X.}~\bibnamefont {Ma}}, \bibinfo {author} {\bibfnamefont
  {Q.}~\bibnamefont {Zhang}},\ and\ \bibinfo {author} {\bibfnamefont {J.-W.}\
  \bibnamefont {Pan}},\ }\bibfield  {title} {\bibinfo {title} {Experimental
  passive round-robin differential phase-shift quantum key distribution},\
  }\href@noop {} {\bibfield  {journal} {\bibinfo  {journal} {Phys. Rev. Lett.}\
  }\textbf {\bibinfo {volume} {114}},\ \bibinfo {pages} {180502} (\bibinfo
  {year} {2015})}\BibitemShut {NoStop}%
\bibitem [{\citenamefont {Li}\ \emph {et~al.}(2016)\citenamefont {Li},
  \citenamefont {Cao}, \citenamefont {Dai}, \citenamefont {Lin}, \citenamefont
  {Zhang}, \citenamefont {Chen}, \citenamefont {Xu}, \citenamefont {Guan},
  \citenamefont {Liao}, \citenamefont {Yin}, \citenamefont {Zhang},
  \citenamefont {Ma}, \citenamefont {Peng},\ and\ \citenamefont
  {Pan}}]{Li2016}%
  \BibitemOpen
  \bibfield  {author} {\bibinfo {author} {\bibfnamefont {Y.-H.}\ \bibnamefont
  {Li}}, \bibinfo {author} {\bibfnamefont {Y.}~\bibnamefont {Cao}}, \bibinfo
  {author} {\bibfnamefont {H.}~\bibnamefont {Dai}}, \bibinfo {author}
  {\bibfnamefont {J.}~\bibnamefont {Lin}}, \bibinfo {author} {\bibfnamefont
  {Z.}~\bibnamefont {Zhang}}, \bibinfo {author} {\bibfnamefont
  {W.}~\bibnamefont {Chen}}, \bibinfo {author} {\bibfnamefont {Y.}~\bibnamefont
  {Xu}}, \bibinfo {author} {\bibfnamefont {J.-Y.}\ \bibnamefont {Guan}},
  \bibinfo {author} {\bibfnamefont {S.-K.}\ \bibnamefont {Liao}}, \bibinfo
  {author} {\bibfnamefont {J.}~\bibnamefont {Yin}}, \bibinfo {author}
  {\bibfnamefont {Q.}~\bibnamefont {Zhang}}, \bibinfo {author} {\bibfnamefont
  {X.}~\bibnamefont {Ma}}, \bibinfo {author} {\bibfnamefont {C.-Z.}\
  \bibnamefont {Peng}},\ and\ \bibinfo {author} {\bibfnamefont {J.-W.}\
  \bibnamefont {Pan}},\ }\bibfield  {title} {\bibinfo {title} {Experimental
  round-robin differential phase-shift quantum key distribution},\ }\href
  {https://doi.org/10.1103/PhysRevA.93.030302} {\bibfield  {journal} {\bibinfo
  {journal} {Phys. Rev. A}\ }\textbf {\bibinfo {volume} {93}},\ \bibinfo
  {pages} {030302} (\bibinfo {year} {2016})}\BibitemShut {NoStop}%
\bibitem [{\citenamefont {Mizutani}\ \emph {et~al.}(2015)\citenamefont
  {Mizutani}, \citenamefont {Imoto},\ and\ \citenamefont
  {Tamaki}}]{Mizutani2015}%
  \BibitemOpen
  \bibfield  {author} {\bibinfo {author} {\bibfnamefont {A.}~\bibnamefont
  {Mizutani}}, \bibinfo {author} {\bibfnamefont {N.}~\bibnamefont {Imoto}},\
  and\ \bibinfo {author} {\bibfnamefont {K.}~\bibnamefont {Tamaki}},\
  }\bibfield  {title} {\bibinfo {title} {Robustness of the round-robin
  differential-phase-shift quantum-key-distribution protocol against source
  flaws},\ }\href@noop {} {\bibfield  {journal} {\bibinfo  {journal} {Phys.
  Rev. A}\ }\textbf {\bibinfo {volume} {92}},\ \bibinfo {pages} {060303}
  (\bibinfo {year} {2015})}\BibitemShut {NoStop}%
\bibitem [{\citenamefont {Hwang}(2003)}]{Hwang2003}%
  \BibitemOpen
  \bibfield  {author} {\bibinfo {author} {\bibfnamefont {W.-Y.}\ \bibnamefont
  {Hwang}},\ }\bibfield  {title} {\bibinfo {title} {Quantum key distribution
  with high loss: Toward global secure communication},\ }\href
  {https://doi.org/10.1103/PhysRevLett.91.057901} {\bibfield  {journal}
  {\bibinfo  {journal} {Phys. Rev. Lett.}\ }\textbf {\bibinfo {volume} {91}},\
  \bibinfo {pages} {057901} (\bibinfo {year} {2003})}\BibitemShut {NoStop}%
\bibitem [{\citenamefont {Yin}\ \emph {et~al.}(2016)\citenamefont {Yin},
  \citenamefont {Fu}, \citenamefont {Mao},\ and\ \citenamefont
  {Chen}}]{Yin2016}%
  \BibitemOpen
  \bibfield  {author} {\bibinfo {author} {\bibfnamefont {H.-L.}\ \bibnamefont
  {Yin}}, \bibinfo {author} {\bibfnamefont {Y.}~\bibnamefont {Fu}}, \bibinfo
  {author} {\bibfnamefont {Y.}~\bibnamefont {Mao}},\ and\ \bibinfo {author}
  {\bibfnamefont {Z.-B.}\ \bibnamefont {Chen}},\ }\bibfield  {title} {\bibinfo
  {title} {Detector-decoy quantum key distribution without monitoring signal
  disturbance},\ }\href@noop {} {\bibfield  {journal} {\bibinfo  {journal}
  {Phys. Rev. A}\ }\textbf {\bibinfo {volume} {93}},\ \bibinfo {pages} {022330}
  (\bibinfo {year} {2016})}\BibitemShut {NoStop}%
\bibitem [{\citenamefont {Zhang}\ \emph {et~al.}(2017)\citenamefont {Zhang},
  \citenamefont {Yuan}, \citenamefont {Cao},\ and\ \citenamefont
  {Ma}}]{Zhang2017}%
  \BibitemOpen
  \bibfield  {author} {\bibinfo {author} {\bibfnamefont {Z.}~\bibnamefont
  {Zhang}}, \bibinfo {author} {\bibfnamefont {X.}~\bibnamefont {Yuan}},
  \bibinfo {author} {\bibfnamefont {Z.}~\bibnamefont {Cao}},\ and\ \bibinfo
  {author} {\bibfnamefont {X.}~\bibnamefont {Ma}},\ }\bibfield  {title}
  {\bibinfo {title} {Practical round-robin differential-phase-shift quantum key
  distribution},\ }\href@noop {} {\bibfield  {journal} {\bibinfo  {journal}
  {New Journal of Physics}\ }\textbf {\bibinfo {volume} {19}},\ \bibinfo
  {pages} {033013} (\bibinfo {year} {2017})}\BibitemShut {NoStop}%
\bibitem [{\citenamefont {Sasaki}\ and\ \citenamefont
  {Koashi}(2017)}]{Sasaki2017}%
  \BibitemOpen
  \bibfield  {author} {\bibinfo {author} {\bibfnamefont {T.}~\bibnamefont
  {Sasaki}}\ and\ \bibinfo {author} {\bibfnamefont {M.}~\bibnamefont
  {Koashi}},\ }\bibfield  {title} {\bibinfo {title} {A security proof of the
  round-robin differential phase shift quantum key distribution protocol based
  on the signal disturbance},\ }\href@noop {} {\bibfield  {journal} {\bibinfo
  {journal} {Quantum Science and Technology}\ }\textbf {\bibinfo {volume}
  {2}},\ \bibinfo {pages} {024006} (\bibinfo {year} {2017})}\BibitemShut
  {NoStop}%
\bibitem [{\citenamefont {Hatakeyama}\ \emph {et~al.}(2017)\citenamefont
  {Hatakeyama}, \citenamefont {Mizutani}, \citenamefont {Kato}, \citenamefont
  {Imoto},\ and\ \citenamefont {Tamaki}}]{Hatakeyama2017}%
  \BibitemOpen
  \bibfield  {author} {\bibinfo {author} {\bibfnamefont {Y.}~\bibnamefont
  {Hatakeyama}}, \bibinfo {author} {\bibfnamefont {A.}~\bibnamefont
  {Mizutani}}, \bibinfo {author} {\bibfnamefont {G.}~\bibnamefont {Kato}},
  \bibinfo {author} {\bibfnamefont {N.}~\bibnamefont {Imoto}},\ and\ \bibinfo
  {author} {\bibfnamefont {K.}~\bibnamefont {Tamaki}},\ }\bibfield  {title}
  {\bibinfo {title} {Differential-phase-shift quantum-key-distribution protocol
  with a small number of random delays},\ }\href@noop {} {\bibfield  {journal}
  {\bibinfo  {journal} {Phys. Rev. A}\ }\textbf {\bibinfo {volume} {95}},\
  \bibinfo {pages} {042301} (\bibinfo {year} {2017})}\BibitemShut {NoStop}%
\bibitem [{\citenamefont {Wang}\ and\ \citenamefont {Zhao}(2017)}]{WangLe2017}%
  \BibitemOpen
  \bibfield  {author} {\bibinfo {author} {\bibfnamefont {L.}~\bibnamefont
  {Wang}}\ and\ \bibinfo {author} {\bibfnamefont {S.}~\bibnamefont {Zhao}},\
  }\bibfield  {title} {\bibinfo {title} {Round-robin differential-phase-shift
  quantum key distribution with heralded pair-coherent sources},\ }\href@noop
  {} {\bibfield  {journal} {\bibinfo  {journal} {Quantum Information
  Processing}\ }\textbf {\bibinfo {volume} {16}},\ \bibinfo {pages} {100}
  (\bibinfo {year} {2017})}\BibitemShut {NoStop}%
\bibitem [{\citenamefont {Liu}\ \emph {et~al.}(2017)\citenamefont {Liu},
  \citenamefont {Guo}, \citenamefont {Qin},\ and\ \citenamefont
  {Wen}}]{Liu2017}%
  \BibitemOpen
  \bibfield  {author} {\bibinfo {author} {\bibfnamefont {L.}~\bibnamefont
  {Liu}}, \bibinfo {author} {\bibfnamefont {F.-Z.}\ \bibnamefont {Guo}},
  \bibinfo {author} {\bibfnamefont {S.-J.}\ \bibnamefont {Qin}},\ and\ \bibinfo
  {author} {\bibfnamefont {Q.-Y.}\ \bibnamefont {Wen}},\ }\bibfield  {title}
  {\bibinfo {title} {Round-robin differential-phase-shift quantum key
  distribution with a passive decoy state method},\ }\href@noop {} {\bibfield
  {journal} {\bibinfo  {journal} {Scientific Reports}\ }\textbf {\bibinfo
  {volume} {7}},\ \bibinfo {pages} {42261} (\bibinfo {year}
  {2017})}\BibitemShut {NoStop}%
\bibitem [{\citenamefont {Leermakers}\ and\ \citenamefont
  {Skoric}(2017)}]{Leermakers2017}%
  \BibitemOpen
  \bibfield  {author} {\bibinfo {author} {\bibfnamefont {D.}~\bibnamefont
  {Leermakers}}\ and\ \bibinfo {author} {\bibfnamefont {B.}~\bibnamefont
  {Skoric}},\ }\bibfield  {title} {\bibinfo {title} {Security proof for round
  robin differential phase shift qkd},\ }\href@noop {} {\bibfield  {journal}
  {\bibinfo  {journal} {arXiv preprint arXiv:1709.00552}\ } (\bibinfo {year}
  {2017})}\BibitemShut {NoStop}%
\bibitem [{\citenamefont {Yin}\ \emph {et~al.}(2018)\citenamefont {Yin},
  \citenamefont {Wang}, \citenamefont {Chen}, \citenamefont {Han},
  \citenamefont {Wang}, \citenamefont {Guo},\ and\ \citenamefont
  {Han}}]{Yin2018}%
  \BibitemOpen
  \bibfield  {author} {\bibinfo {author} {\bibfnamefont {Z.-Q.}\ \bibnamefont
  {Yin}}, \bibinfo {author} {\bibfnamefont {S.}~\bibnamefont {Wang}}, \bibinfo
  {author} {\bibfnamefont {W.}~\bibnamefont {Chen}}, \bibinfo {author}
  {\bibfnamefont {Y.-G.}\ \bibnamefont {Han}}, \bibinfo {author} {\bibfnamefont
  {R.}~\bibnamefont {Wang}}, \bibinfo {author} {\bibfnamefont {G.-C.}\
  \bibnamefont {Guo}},\ and\ \bibinfo {author} {\bibfnamefont {Z.-F.}\
  \bibnamefont {Han}},\ }\bibfield  {title} {\bibinfo {title} {Improved
  security bound for the round-robin-differential-phase-shift quantum key
  distribution},\ }\href@noop {} {\bibfield  {journal} {\bibinfo  {journal}
  {Nature communications}\ }\textbf {\bibinfo {volume} {9}},\ \bibinfo {pages}
  {457} (\bibinfo {year} {2018})}\BibitemShut {NoStop}%
\bibitem [{\citenamefont {Bouchard}\ \emph {et~al.}(2018)\citenamefont
  {Bouchard}, \citenamefont {Sit}, \citenamefont {Heshami}, \citenamefont
  {Fickler},\ and\ \citenamefont {Karimi}}]{Bouchard2018}%
  \BibitemOpen
  \bibfield  {author} {\bibinfo {author} {\bibfnamefont {F.}~\bibnamefont
  {Bouchard}}, \bibinfo {author} {\bibfnamefont {A.}~\bibnamefont {Sit}},
  \bibinfo {author} {\bibfnamefont {K.}~\bibnamefont {Heshami}}, \bibinfo
  {author} {\bibfnamefont {R.}~\bibnamefont {Fickler}},\ and\ \bibinfo {author}
  {\bibfnamefont {E.}~\bibnamefont {Karimi}},\ }\bibfield  {title} {\bibinfo
  {title} {Round-robin differential-phase-shift quantum key distribution with
  twisted photons},\ }\href {https://doi.org/10.1103/PhysRevA.98.010301}
  {\bibfield  {journal} {\bibinfo  {journal} {Phys. Rev. A}\ }\textbf {\bibinfo
  {volume} {98}},\ \bibinfo {pages} {010301} (\bibinfo {year}
  {2018})}\BibitemShut {NoStop}%
\bibitem [{\citenamefont {Kawakami}\ \emph {et~al.}(2016)\citenamefont
  {Kawakami}, \citenamefont {Sasaki},\ and\ \citenamefont
  {Koashi}}]{Kawakami2016}%
  \BibitemOpen
  \bibfield  {author} {\bibinfo {author} {\bibfnamefont {S.}~\bibnamefont
  {Kawakami}}, \bibinfo {author} {\bibfnamefont {T.}~\bibnamefont {Sasaki}},\
  and\ \bibinfo {author} {\bibfnamefont {M.}~\bibnamefont {Koashi}},\
  }\bibfield  {title} {\bibinfo {title} {Security of the
  differential-quadrature-phase-shift quantum key distribution},\ }\href@noop
  {} {\bibfield  {journal} {\bibinfo  {journal} {Phys. Rev. A}\ }\textbf
  {\bibinfo {volume} {94}},\ \bibinfo {pages} {022332} (\bibinfo {year}
  {2016})}\BibitemShut {NoStop}%
\bibitem [{\citenamefont {Koashi}(2009)}]{Koashi2009}%
  \BibitemOpen
  \bibfield  {author} {\bibinfo {author} {\bibfnamefont {M.}~\bibnamefont
  {Koashi}},\ }\bibfield  {title} {\bibinfo {title} {Simple security proof of
  quantum key distribution based on complementarity},\ }\href@noop {}
  {\bibfield  {journal} {\bibinfo  {journal} {New Journal of Physics}\ }\textbf
  {\bibinfo {volume} {11}},\ \bibinfo {pages} {045018} (\bibinfo {year}
  {2009})}\BibitemShut {NoStop}%
\bibitem [{\citenamefont {Hayashi}\ and\ \citenamefont
  {Tsurumaru}(2012)}]{Hayashi2012}%
  \BibitemOpen
  \bibfield  {author} {\bibinfo {author} {\bibfnamefont {M.}~\bibnamefont
  {Hayashi}}\ and\ \bibinfo {author} {\bibfnamefont {T.}~\bibnamefont
  {Tsurumaru}},\ }\bibfield  {title} {\bibinfo {title} {Concise and tight
  security analysis of the bennett^^e2^^80^^93brassard 1984 protocol with
  finite key lengths},\ }\href@noop {} {\bibfield  {journal} {\bibinfo
  {journal} {New Journal of Physics}\ }\textbf {\bibinfo {volume} {14}},\
  \bibinfo {pages} {093014} (\bibinfo {year} {2012})}\BibitemShut {NoStop}%
\bibitem [{\citenamefont {Tsurumaru}\ and\ \citenamefont
  {Hayashi}(2013)}]{Tsurumaru2013}%
  \BibitemOpen
  \bibfield  {author} {\bibinfo {author} {\bibfnamefont {T.}~\bibnamefont
  {Tsurumaru}}\ and\ \bibinfo {author} {\bibfnamefont {M.}~\bibnamefont
  {Hayashi}},\ }\bibfield  {title} {\bibinfo {title} {Dual universality of hash
  functions and its applications to quantum cryptography},\ }\href@noop {}
  {\bibfield  {journal} {\bibinfo  {journal} {IEEE Trans. Inf. Theory}\
  }\textbf {\bibinfo {volume} {59}},\ \bibinfo {pages} {4700} (\bibinfo {year}
  {2013})}\BibitemShut {NoStop}%
\bibitem [{\citenamefont {Lim}\ \emph {et~al.}(2014)\citenamefont {Lim},
  \citenamefont {Curty}, \citenamefont {Walenta}, \citenamefont {Xu},\ and\
  \citenamefont {Zbinden}}]{Lim2014}%
  \BibitemOpen
  \bibfield  {author} {\bibinfo {author} {\bibfnamefont {C.~C.~W.}\
  \bibnamefont {Lim}}, \bibinfo {author} {\bibfnamefont {M.}~\bibnamefont
  {Curty}}, \bibinfo {author} {\bibfnamefont {N.}~\bibnamefont {Walenta}},
  \bibinfo {author} {\bibfnamefont {F.}~\bibnamefont {Xu}},\ and\ \bibinfo
  {author} {\bibfnamefont {H.}~\bibnamefont {Zbinden}},\ }\bibfield  {title}
  {\bibinfo {title} {Concise security bounds for practical decoy-state quantum
  key distribution},\ }\href@noop {} {\bibfield  {journal} {\bibinfo  {journal}
  {Phys. Rev. A}\ }\textbf {\bibinfo {volume} {89}},\ \bibinfo {pages} {022307}
  (\bibinfo {year} {2014})}\BibitemShut {NoStop}%
\bibitem [{\citenamefont {Gottesman}\ \emph {et~al.}(2004)\citenamefont
  {Gottesman}, \citenamefont {Lo}, \citenamefont {Lutkenhaus},\ and\
  \citenamefont {Preskill}}]{Gottesman2004}%
  \BibitemOpen
  \bibfield  {author} {\bibinfo {author} {\bibfnamefont {D.}~\bibnamefont
  {Gottesman}}, \bibinfo {author} {\bibfnamefont {H.-K.}\ \bibnamefont {Lo}},
  \bibinfo {author} {\bibfnamefont {N.}~\bibnamefont {Lutkenhaus}},\ and\
  \bibinfo {author} {\bibfnamefont {J.}~\bibnamefont {Preskill}},\ }\bibfield
  {title} {\bibinfo {title} {Security of quantum key distribution with
  imperfect devices},\ }in\ \href@noop {} {\emph {\bibinfo {booktitle}
  {Information Theory, 2004. ISIT 2004. Proceedings. International Symposium
  on}}}\ (\bibinfo  {publisher} {IEEE},\ \bibinfo {year} {2004})\ p.\ \bibinfo
  {pages} {136}\BibitemShut {NoStop}%
\bibitem [{\citenamefont {Csisz\'{a}r}(1984)}]{Csiszar1984}%
  \BibitemOpen
  \bibfield  {author} {\bibinfo {author} {\bibfnamefont {I.}~\bibnamefont
  {Csisz\'{a}r}},\ }\bibfield  {title} {\bibinfo {title} {Sanov property,
  generalized i-projection and a conditional limit theorem},\ }\href@noop {}
  {\bibfield  {journal} {\bibinfo  {journal} {The Annals of Probability}\
  }\textbf {\bibinfo {volume} {12}},\ \bibinfo {pages} {768} (\bibinfo {year}
  {1984})}\BibitemShut {NoStop}%
\bibitem [{\citenamefont {Sanov}(1957)}]{Sanov1957}%
  \BibitemOpen
  \bibfield  {author} {\bibinfo {author} {\bibfnamefont {I.~N.}\ \bibnamefont
  {Sanov}},\ }\bibfield  {title} {\bibinfo {title} {On the probability of large
  deviations of random magnitudes},\ }\href@noop {} {\bibfield  {journal}
  {\bibinfo  {journal} {Mat. Sb. (NS),}\ }\textbf {\bibinfo {volume}
  {42(84)}},\ \bibinfo {pages} {11} (\bibinfo {year} {1957})}\BibitemShut
  {NoStop}%
\bibitem [{\citenamefont {Cover}\ and\ \citenamefont
  {Thomas}(2012)}]{Cover2012}%
  \BibitemOpen
  \bibfield  {author} {\bibinfo {author} {\bibfnamefont {T.~M.}\ \bibnamefont
  {Cover}}\ and\ \bibinfo {author} {\bibfnamefont {J.~A.}\ \bibnamefont
  {Thomas}},\ }\href@noop {} {\emph {\bibinfo {title} {Elements of information
  theory}}}\ (\bibinfo  {publisher} {John Wiley \& Sons},\ \bibinfo {year}
  {2012})\BibitemShut {NoStop}%
\bibitem [{\citenamefont {Bernstein}(1924)}]{Bernstein1924}%
  \BibitemOpen
  \bibfield  {author} {\bibinfo {author} {\bibfnamefont {S.}~\bibnamefont
  {Bernstein}},\ }\bibfield  {title} {\bibinfo {title} {On a modification of
  chebyshev’s inequality and of the error formula of laplace},\ }\href@noop {}
  {\bibfield  {journal} {\bibinfo  {journal} {Ann. Sci. Inst. Sav. Ukraine,
  Sect. Math}\ }\textbf {\bibinfo {volume} {1}},\ \bibinfo {pages} {38}
  (\bibinfo {year} {1924})}\BibitemShut {NoStop}%
\end{thebibliography}%

\end{document}